\documentclass{article}

\usepackage{arxiv}

\usepackage[utf8]{inputenc} 
\usepackage[T1]{fontenc}    
\usepackage{hyperref}       
\usepackage{url}            
\usepackage{booktabs}       
\usepackage{amsfonts}       
\usepackage{nicefrac}       
\usepackage{microtype}      
\usepackage{lipsum}
\usepackage{fancyhdr}       
\usepackage{graphicx}       
\graphicspath{{media/}}     

\usepackage{amssymb}

\usepackage{amsmath}

\usepackage{booktabs}
\usepackage{booktabs}
\usepackage{algorithm}
\usepackage{algorithmic}
\usepackage{subcaption}
\usepackage{orcidlink}

\title{DeepTrust: Multi-Step Classification through Dissimilar Adversarial Representations for Robust Android Malware Detection}

\author{
  Daniel Pulido-Cortázar, \space Daniel Gibert, \space Felip Manyà \\
  Artificial Intelligence Research Institute (IIIA-CSIC) \\
  Bellaterra, Spain\\
  \texttt{danielpulidocortazar@gmail.com} \\
  \texttt{\{daniel.gibert,felip\}@iiia.csic.es} \\
}

\begin{document}
\maketitle

\begin{abstract}
Over the last decade, machine learning has been extensively applied to identify malicious Android applications. However, such approaches remain vulnerable against adversarial examples, i.e., examples that are subtly manipulated to fool a machine learning model into making incorrect predictions.
This research presents DeepTrust, a novel metaheuristic that arranges flexible classifiers, like deep neural networks, into an ordered sequence where the final decision is made by a single internal model based on conditions activated in cascade. In the Robust Android Malware Detection competition at the 2025 IEEE Conference SaTML, DeepTrust secured the first place and achieved state-of-the-art results, outperforming the next-best competitor by up to 266\% under feature-space evasion attacks. This is accomplished while maintaining the highest detection rate on non-adversarial malware and a false positive rate below 1\%. The method's efficacy stems from maximizing the divergence of the learned representations among the internal models. By using classifiers inducing fundamentally dissimilar embeddings of the data, the decision space becomes unpredictable for an attacker. This frustrates the iterative perturbation process inherent to evasion attacks, enhancing system robustness without compromising accuracy on clean examples.
\end{abstract}


\section{Introduction}

The Android operating system (OS) has seen rapid development, becoming the most popular OS for smart mobile devices since its release in 2008. By the first quarter of 2025, Android-based smartphones accounted for approximately 80\% of global sales \cite{android-share}, and in June 2025, Google Play, the official store for Android applications, is hosting over 1,5 million applications \cite{android-apps}. The open ecosystem of Android, its coarse-grained permission management, and the ability to invoke third-party code, create numerous security attack surfaces~\cite{liu_review_2020}. Traditional antivirus methods, such as fingerprinting and blacklisting, have proven insufficient in safeguarding mobile users, especially against encrypted and zero-day malware. This has prompted researchers in the Android malware domain to explore solutions based on machine learning (ML) \cite{liu_review_2020,DBLP:conf/ndss/ArpSHGR14, 10.1109/TDSC.2017.2700270,qiu_survey_2021,guerra-manzanares_machine_2024}. In such a setting, collected Android applications are generally processed to encode their information in highly dimensional feature vectors. These vectors are then used to train ML algorithms, such as Support Vector Machines (SVMs) and deep neural networks, aiming to identify malicious applications end-to-end. 

Despite advancements, recent works \cite{guerra-manzanares_machine_2024,pendlebury_tesseract_2019,molina-coronado_towards_2023} have highlighted persistent limitations in current methodologies such as lack of reproducibility, poor dataset quality and inadequate evaluation metrics. These limitations often lead to an overestimation of the performance of detectors in research settings, further exacerbated by the lack of standardized benchmark datasets that lead to inconsistent evaluations and unfair comparisons. Moreover, current evaluation procedures overlook the inherent adversarial nature of the Android malware domain, where attackers are economically motivated to evade detection. Furthermore, evaluations rarely consider data distribution drift, where applications evolve over time, causing the trained models to degrade in effectiveness \cite{pendlebury_tesseract_2019}. As a result, Android malware detectors are commonly assessed in steady, idealized conditions that fail to represent real-world adversarial conditions.

Aiming to address these shortcomings, the European Lighthouse on Secure and Safe AI (ELSA)\footnote{\url{https://elsa-ai.eu/}} has proposed the Robust Android Malware Detection Benchmark (ELSA-RAMD) \cite{ramd-elsa-benchmark}, organized as an official competition \cite{ramd-elsa-competition} at the 2025 IEEE Conference on Secure and Trustworthy Machine Learning (SaTML'25).\footnote{\url{https://satml.org/2025/}} The benchmark provides a standardized and open framework for evaluating static, learning based Android malware detectors by assessing robustness to feature-space attacks, problem-space attacks, and temporal data drift. It enforces lower-bound performance constraints, such as a low False Positive Rate ($\leq1\%$), and uses functionality-preserving attacks to manipulate malware. 

The primary contribution of this paper is DeepTrust\footnote{\url{https://github.com/danielPulidoCortazar/deeptrust}}, a particular realization of a novel multi-step defensive metaheuristic that arranges flexible classifiers, such as deep neural networks, in an ordered sequence. In this framework, an input is evaluated sequentially, and a final decision is rendered by a single constituent model based on a cascade of activation conditions, rather than by an aggregation of outputs as in conventional ensemble methods. The central hypothesis, empirically studied in this work, is that the efficacy of multi-step classification is contingent upon maximizing the divergence of the learned representations among the internal models. By employing classifiers projecting the input to dissimilar dense representations, the decision space becomes unpredictable for an attacker. This frustrates the iterative perturbation process inherent to evasion attacks and enhances system robustness without compromising classification accuracy on non-adversarial examples. 

DeepTrust, a particular configuration of our proposal, achieves state-of-the-art results and was awarded the gold medal in the ELSA-RAMD benchmark. Concretely, it surpass previous state-of-the-art baselines and the next-best competitor by a relative margin of up to 266\% under feature-space evasion attack and up to 217\% under problem-space evasion attack while simultaneously maintaining the highest detection rate on non-adversarially perturbed malware and a False Positive Rate under 1\%.

Section \ref{sec:related-work} reviews the current literature relevant to our domain. Section \ref{sec:threat-model} details the ELSA-RAMD threat model framework, which serves as the foundation of our evaluation methodology. Section \ref{sec:deeptrust} introduces the multi-step architecture underlying DeepTrust. Section \ref{sec:evaluation} presents our experimental setup and discusses the results obtained on the ELSA-RAMD benchmark. Finally, Section \ref{sec:conclusions} summarizes our findings, outlines limitations, and suggests directions for future work.

\section{Related Work}\label{sec:related-work}

ML approaches to malware detection are generally categorized into static and dynamic detection \cite{liu_review_2020}.
Static detection involves examining the code of an Android application without executing it. This method offers high code coverage. However, static analysis is challenged from countermeasures like code obfuscation and dynamic code loading \cite{qiu_survey_2021, MOLINACORONADO2025104094}. Conversely, dynamic detection records the behavior of an Android application during execution. This approach reveals the actual runtime behavior and it is generally more robust against obfuscation and dynamic loading. Despite the benefits, dynamic detection requires significant computational resources and time, becoming impractical in most use cases.

We center our focus on static analysis proposals, the most prolific category for learning-based detectors, and where our method emerges as a robust candidate. The primary object of static analysis is the Android Application Package (APK) file, which can be decompiled to extract files like AndroidManifest.xml and smali files. In this context, the DREBIN framework \cite{DBLP:conf/ndss/ArpSHGR14} represents a foundational and widely recognized methodology for Android malware detection, initially proposed in 2014. The methodology involves extracting a comprehensive set of features directly from APKs.

The development of learning-based methods for malware detection in the Android ecosystem has driven the research community to build open-access datasets. We identify the main open datasets available to be Drebin \cite{DBLP:conf/ndss/ArpSHGR14}, MalGenome \cite{zhou_dissecting_2012}, AMD \cite{polychronakis_deep_2017}, and RmvDroid \cite{wang_rmvdroid_2019}. Static datasets can become quickly outdated as applications, technologies and the Android operating system itself evolve. The initiative AndroZoo \cite{allix_androzoo_2016} tries to overcome this issue by providing a living repository of Android applications with rich metadata including size, hash values, the list of requested permissions from the Androidmanifest.xml file, and reports fromVirusTotal.\footnote{\url{https://www.virustotal.com}}

Based on recent surveys \cite{liu_review_2020, qiu_survey_2021, gibert_rise_2020}, learning-based algorithms are varied and in general lie within linear models, naive Bayes, ensembling, support vector machines (SVM) and deep learning. The evaluation methodology in the literature primarily consists of evaluating on a held-out test partition and use common metrics for binary classification problems such as Accuracy, Precision, Recall or True Positive Rate, Specificity or True Negative Rate, F1 score and area under the Receiver-operating characteristic curve (AUC). There is no well principled evaluation under evasion attacks or in the wild, as pointed out in \cite{guerra-manzanares_machine_2024, pendlebury_tesseract_2019,molina-coronado_towards_2023}.

We evaluate and compare DeepTrust through the SaTML’25 competition, where participants submitted well-established methods from the literature. ELSA-RAMD, the open benchmark serving as framework by the competition, sets a realistic and fair evaluation paradigm to assess alternatives under different common assumptions, making it the first proposed benchmark in the problem domain of Android malware detection. The proposals submitted to the competition (i.e., achieved a false positive rate under 1\%) are DREBIN \cite{DBLP:conf/ndss/ArpSHGR14}, SecureSVM \cite{10.1109/TDSC.2017.2700270}, Support Vector Machine with Custom Bounds (SMV-CB) \cite{DBLP:conf/itasec/AngioniDPB22}, Continual-Positive Congruent Training (C-PCT) \cite{ghiani2025regressionawarecontinuallearningandroid} and DeepTrust. In addition to the participants, we expand the benchmark by including other competing ML algorithms for tabular data, such as Random Forest \cite{tin_kam_ho_random_1995}, XGBoost \cite{10.1145/2939672.2939785} and robust feed-forward neural networks adversarially trained with the algorithm proposed in Section \ref{sec:tab-adv}.

DREBIN \cite{DBLP:conf/ndss/ArpSHGR14}, the most popular framework for malware detection in Android, begins by statically inspecting a given APK and extracting a comprehensive set of features from its manifest file (AndroidManifest.xml) and disassembled Dalvik bytecode (dex code). Then an SVM determines a hyperplane that optimally separates the two classes, malware and goodware, in the vector space. The DREBIN feature extractor is adopted in Track 1 of ELSA-RAMD, where it is simulated a feature-space attack in which the adversary or attacker knows the features used by the targeted ML system.  Subsequently, the work done by A. Demontis et al. \cite{10.1109/TDSC.2017.2700270} studied DREBIN vulnerabilities to evasion attacks. The paper specifically shows that DREBIN's performance can be ``significantly downgraded in the presence of skilled attackers that can carefully manipulate malware samples to evade classifier detection"\cite{10.1109/TDSC.2017.2700270}. The paper's core contribution is Sec-SVM, a novel, theoretically-sound learning algorithm to train linear classifiers with more evenly distributed feature weights that harden the sensitivity of algorithm outputs to slight feature changes. From a different perspective, authors in \cite{DBLP:conf/itasec/AngioniDPB22} analyze the performance degradation of DREBIN due to the natural evolution of malware over time. Leveraging the information obtained from their ``drift-analysis framework", the paper proposes SVM-CB, a classifier designed to clip weights of unstable features over time, measured by means of a novel metric called Temporal Feature Stability. Lastly, C-PCT, developed in \cite{ghiani2025regressionawarecontinuallearningandroid}, is a strategy that departs from \cite{9577424} for Continual Learning scenarios. In this setting, a Multi-layer Perceptron (MLP) is trained with data introduced sequentially over time, and the model must learn from new information without forgetting the old. C-PCT addresses this by encouraging the updated model to behave like the previous version, but with a key exception: it only imitates the prior model's behavior on predictions that were correct.

\section{Threat Model}\label{sec:threat-model}
In this work, we adopt a threat model aligned with the constraints of the ELSA-RAMD benchmark and the rules specified in its competition hosted at IEEE SaTML'25. ELSA-RAMD is structured in three separate tracks (a complete description of each track is provided in Section \ref{sec:elsa-cyber}) that evaluate the robustness of machine learning-based Android malware detectors against feature-space attacks (Track 1), problem-space attacks (Track 2) and data drift (Track 3).

The threat model in this work focuses primarily on the worst-case evasion attack scenario simulated in \emph{Track 1: Adversarial Robustness to Feature-space Attacks}. In this track, the performance of the detection system is measured against increasing amounts of adversarial manipulations, assuming the attacker has knowledge of the features used, i.e., the DREBIN features \cite{DBLP:conf/ndss/ArpSHGR14}, the transition from raw model prediction to class attribution and has unlimited access to model usage. 

\subsection{DREBIN features}
\label{sec:drebin_features}
The DREBIN features \cite{DBLP:conf/ndss/ArpSHGR14} are a set of lightweight, statically extracted features from APKs. Each APK is represented as a high-dimensional, sparse binary feature vector, $x \in \{ 0,1\}^d$, where $d$ equals to $1,461,078$ tokens in the ELSA-RAMD training dataset. Tokens capture both syntactic and semantic information extracted from two main sources:
\begin{enumerate}
    \item AndroidManifest.xml.
    \item Disassembled code (Dex code / classes.dex).
\end{enumerate}
The following feature sets are extracted from the AndroidManifest.xml:
\begin{enumerate}
    \item Hardware Components (S1).
    \item Requested Permissions (S2).
    \item App Components (S3).
    \item Filtered Intents (S4).
\end{enumerate}
The following feature sets are extracted from the disassembled code:
\begin{enumerate}
    \item Restricted API Calls (S5).
    \item Used Permissions (S6).
    \item Suspicious API Calls (S7).
    \item Network Addresses (S8).
\end{enumerate}
The joint feature set $S$ can be represented as
$$ S:= S_1 \cup S_2 \cup ... \cup S_8,$$
where $|S| = d $.

\subsection{Attacker Capabilities}
Within the scope of ELSA-RAMD Track 1, the attacker operates in the feature-space, i.e., the input feature vector, $x \in \{0,1\}^d$, is perturbed, abstracting away the implementation details at the APK level. \footnote{This abstraction, while useful for studying the theoretical limits of model robustness, introduces a potential gap between feature-space attacks and problem-space attacks. Thus, although an adversarial feature vector $x'$ may successfully evade the detection model in feature-space, it does not guarantee that a corresponding functional APK can be constructed to realize that vector. To this end, the ELSA-RAMD benchmark evaluates on a different track (Track 2) the robustness of the models to realizable, real-world attacks, where the actual APKs are modified instead of the input feature vectors.} 

Therefore, we assume a gray-box threat model, under which the attacker has:
\begin{itemize}
    \item Full knowledge of the feature extraction process, including the DREBIN feature representation of an APK.
    \item Limited knowledge of the model architecture: the semantics of the prediction output and how these predictions are used to attribute a label to the input.
    \item Unlimited query access to the detection system, allowing the attacker to observe the classification outcomes and confidence scores for every input.
\end{itemize}
Given these assumptions, the attacker is capable of crafting adversarial examples
$x' \in \{0, 1\}^d$ by modifying at most $k$ features in the original feature vector $x$, i.e., $|| x' -x ||_{l_1} \leq k$. The goal of the attacker is to preserve the malicious functionality of the APK while inducing the classifier to classify $x'$ as benign.

\subsection{Defender Capabilities}
The objective of the defender is to design a malware detection system that remains effective under adversarial conditions while satisfying the following operational constraints defined by the IEEE SaTML'25 competition.

\subsubsection{Feature Constraints}
The defender is restricted to using only the DREBIN feature set, or a subset thereof, for both training and development. That is, each Android application is represented as a binary vector $x \in \{0,1\}^d$, where $d\leq1,461,078$. No additional sources of information (e.g., dynamic analysis, network traffic, or behavioral logs) are available to the defender.

\subsubsection{Evaluation Framework}
The malware detection system is evaluated using ELSA-RAMD's standardised protocol \cite{ramd-elsa-benchmark}, which includes both clean and adversarial samples generated with feature-space (cf. \ref{sec:feature_space_attack}) and problem-space attacks (cf. \ref{sec:problem_space_attack}). To qualify, submitted methods are required to keep the false positive rate below 1\% on benign data. This constraint reflects the operational requirement that benign applications must rarely be misclassified as malicious, which is critical in real-world deployments where false alarms can lead to user dissatisfaction, degraded system performance, or unnecessary app blocking.

\section{Framework} 
\label{sec:deeptrust}

In this section, we introduce DeepTrust, a multi-step classification framework designed for Android malware detection in accordance with the IEEE SaTML'25 competition rules, i.e., the system encodes an application as a static set of features and can be enforced to not surpass a strict operating point of
1\% false positive rate (FPR).

\begin{figure}[h!]
    \centering
    \includegraphics[width=0.35\columnwidth]{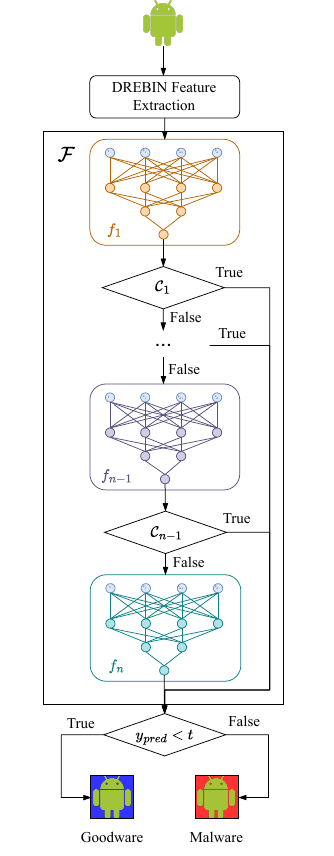}
    \caption{Representation of DeepTrust multi-step architecture for a sequence of length $n$ with different internal models. At decision time, a threshold is used to classify using the system's output.}
    \label{fig:deeptrust}
\end{figure}

Unlike traditional ensemble methods that aggregate predictions across multiple classifiers (e.g., Random Forests \cite{tin_kam_ho_random_1995} or Gradient Boosting \cite{friedman_greedy_2001}), DeepTrust adopts a metaheuristic approach that arranges its internal detectors in a fixed sequential order. At each step, a detector independently predicts the output and a boolean condition is evaluated. If the condition is met or activated, the process terminates early with the detector's prediction being the system's final output. Otherwise, the sample is passed to the next stage. If no condition is met, the last internal model in the sequence is used to render the final prediction. 
The proposed architecture is illustrated in Figure~\ref{fig:deeptrust}.

Traditional ensemble learning methods have been shown to be highly effective for regression and classification tasks involving tabular data \cite{SHWARTZZIV202284}. However, under adversarial conditions, the aggregation mechanism of these methods can become a liability. This is because an adversary with access to the prediction confidence based on the fusion of multiple weak detectors can iteratively craft perturbations that gradually shift the ensemble's final decision. As ensemble outputs are formed by combining many partially accurate predictions, attackers only need to navigate a unique decision space to change the overall outcome, without necessarily needing to fool each one individually or a subset thereof (cf.~Figure \ref{fig:visual-example-whit-box-attacks}, panel A). In contrast, DeepTrust's sequential decision-making requires an adversary to simultaneously deceive multiple and diverse projected representations of the problem learned by the internal models. Because predictions are not aggregated, the system offers better adversarial resilience, as it becomes significantly harder to simultaneously deceive all detectors in the sequence, provided each detector is flexible and strong enough to be used individually, such as deep neural networks

\subsection{Multi-Step Classification through Diverse Representations}\label{sec:multi-step-theory}
The core intuition behind DeepTrust is to build a multi-step sequential classification framework using models encoding diverse dense representations of the data. This framework is agnostic to the number of internal classifiers and defined conditions.
Consider a tabular binary classification task, exemplified with the malware detection problem modeled as described in Section \ref{sec:threat-model}. Let each sample, $x \in \mathbb{R}^d$, be associated with a binary label, $y \in \{0,1\}$, and drawn independently from the same data distribution $\mathcal{D}$. We define an ordered set of $n$ parametrized classifiers, $F:= \{f_1, f_2, ...,f_n\}$, such that $f_i: \mathbb{R}^d \rightarrow [0,1]$ for all $i = 1,2,...,n$, i.e., each classifier produces a confidence prediction conditioned to the input feature values. Let us define an ordered set of activation conditions, $C$, such that $C:= \{\mathcal{C}_1, \mathcal{C}_2, ..., \mathcal{C}_{n-1}\}$. In these terms, multi-step classification can be defined as a tuple $\mathcal{F}:=(F,C,t = \alpha\in \mathbb{R})$. The metaheuristic, sequentially evaluates activation conditions until one is met, and then uses the corresponding classifier to render the final system's prediction. For a given data point, $x$, the multi-step classifier $\mathcal{F}$ will proceed as follows:

\begin{equation}
\mathcal{F}(x) :=
\begin{cases}
    f_1(x) & \text{if } \mathcal{C}_1(x) \text{ is true}, \\
    f_{2}(x) & \text{else if } \mathcal{C}_{2}(x) \text{ is true}, \\
    \vdots & \vdots \\
    f_{n-1}(x) & \text{else if } \mathcal{C}_{n-1}(x) \text{ is true}, \\
    f_n(x) & \text{otherwise}. \\
\end{cases}\ \ \ 
\label{eq:multi-step}
\end{equation}

\begin{figure*}[ht]
    \centering
    \includegraphics[width=\textwidth]{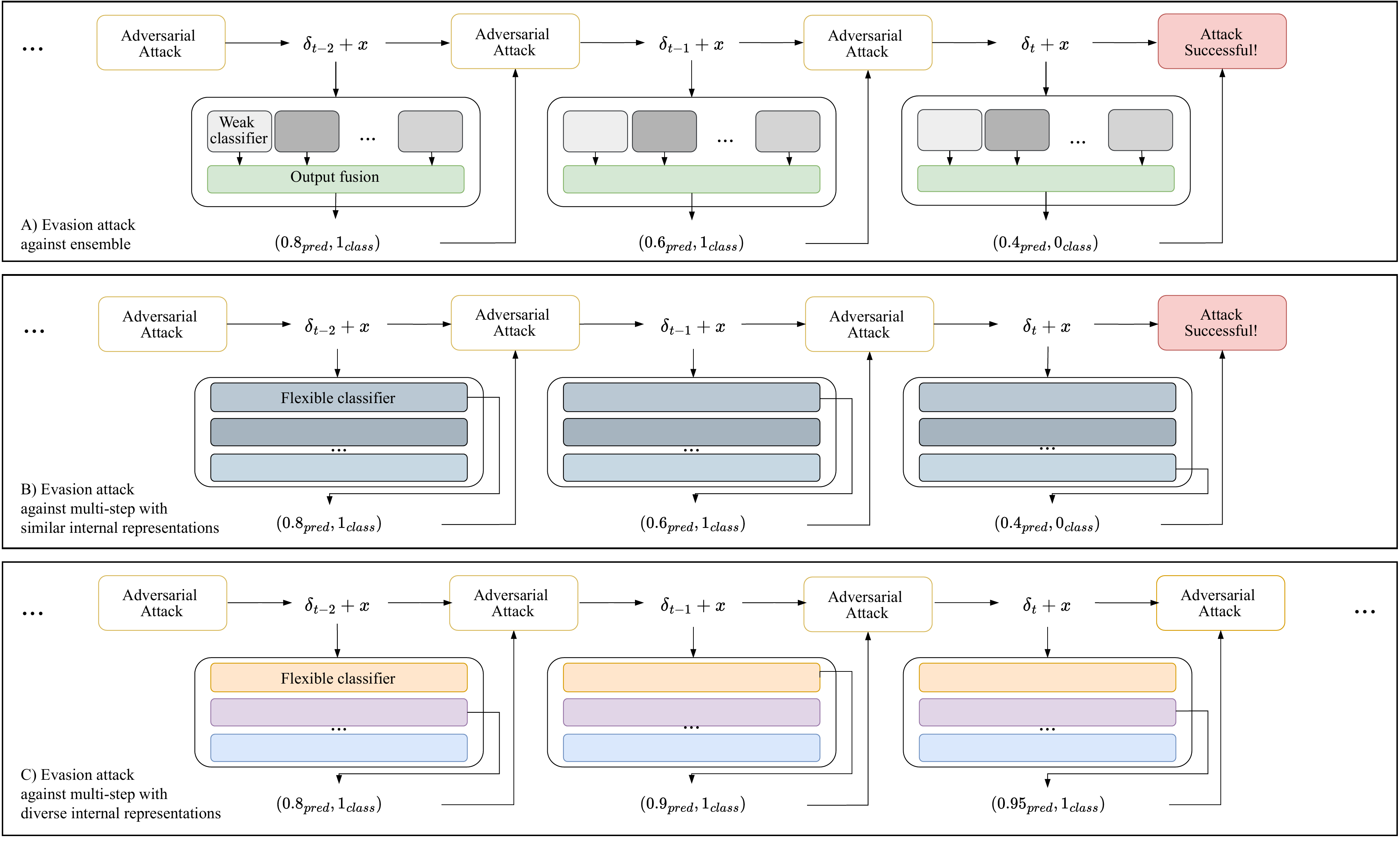}
    \caption{Hypothetical behavior of different metaheuristics under a feature-space evasion attack in a binary classification task. A) Ensemble. An adversary iteratively crafts the appropriate perturbation so the predicted output is the negative (benign) class. As it has access to probabilities and output class, it can iteratively modify their adversarial example until it is misclassified by reducing the predicted probability. B) Multi-step classification using models with similar learned representations. In a similar fashion, having sequential classification performed by powerful models, but with similar internal representations of the data, makes no impact to robustness, as once it gets to deceive the first internal model, the others will probably misclassify the adversarial example as well. C) Multi-step classification using models with diverse learned representations. In a multi-step classification system that uses powerful models with diverse embedding representations, an adversary will attempt to iteratively cause misclassifications. Lacking information about which model's output is being used for the final prediction, the adversary must prepare a new perturbation for the next model. This new perturbation will likely begin to move away from how a benign class is represented internally by the models that were already bypassed, hardening the iterative search for a successful perturbation.}
    \label{fig:visual-example-whit-box-attacks}
\end{figure*}

At classification time, $x$ is labeled as goodware if $\mathcal{F}(x) < t$. Otherwise, $x$ is classified as malicious. Activation condition, $\mathcal{C}_i(x)$, denotes a stage-specific condition that determines whether to classify the input in stage $i$ or to pass the classification decision to the next detector in stage $i+1$. This condition could depend on $f_i$ (e.g, on a confidence score exceeding a threshold $\sigma_i \in [0,1]$), the detection of an anomaly (e.g., via an Isolation Forest \cite{liu_isolation-based_2012} algorithm), or any other criterion. If the condition  $\mathcal{C}_i(x)$ is satisfied, the classifier issues a final prediction. Otherwise, the input is passed to the next detector in the sequence. If none of the intermediate classifiers meet their activation condition, the prediction is made by the last classifier $f_n$. In this framework, if detectors in $F$ encode similar dense representations of the data, adversarial examples that fool one model are likely to fool the others as well, reducing the effectiveness of the sequential defense (cf. Figure \ref{fig:visual-example-whit-box-attacks}, panel B). To avoid this, we explicitly encourage diversity in the internal representations of the detectors by training them with distinct schemes. Concretely, we introduce an adversarial training algorithm adapted to tabular binary data and a label smoothing method. We empirically show in Section \ref{sec:multi-step} that prefixing a specific model architecture (MLP) and training it through different robust and non-robust strategies is a simple and fast-forward approach to get models that encode dissimilar data embeddings for the multi-step framework.

Intuitively, distinct learning algorithms for each model can help them converge to parameters encoding different decision boundaries and feature sensitivities, forcing attackers to satisfy multiple, potentially orthogonal, constraints to succeed. As hypothesized in Figure~\ref{fig:visual-example-whit-box-attacks}, panel C, this can lead to adversarial examples that either fail to pass through all stages or become easily detectable due to the unnatural perturbations applied. Moreover, since the models operate internally and only one of them generates the final output, an attacker cannot easily infer which model was responsible for the classification, further complicating any attack strategy. This unpredictability, combined with divergent internal representations, offers a novel path for improving adversarial robustness without sacrificing classification performance, provided each internal detector is a strong learner.

\subsection{Building Detectors with Diverse Internal Representations} \label{sec:getting-divergence}
To build detectors that, although optimized for the same task (i.e., Android malware detection using the DREBIN feature representation), learn distinct dense representations of the data, in this work we have explored two strategies:
\begin{itemize}
    \item Adversarial training, which builds detectors resilient to manipulations by exposing them to perturbed examples, $x+\delta \in \mathbb{R}^d$, during training.
    \item Label smoothing, which reduces overconfidence by replacing discrete true labels, $y \in \{0,1\}$, with smoothed versions, $y_{s} \in [0,1]$.
\end{itemize}

\subsubsection{Adversarial Training on Tabular Data}\label{sec:tab-adv}
Adversarial training \cite{DBLP:journals/corr/GoodfellowSS14} is a well-known defensive technique used to improve the robustness of machine learning-based models against adversarial attacks. Adversarial training involves generating adversarial examples during the learning phase and using these examples instead of, or in addition to, clean inputs to update the model. This requires multiple forward and backward passes per optimization step, making it more expensive than standard techniques. To speed-up training, in this work we adapt the adversarial training method presented by Shafahi et al. \cite{10.5555/3454287.3454589} to the nuances of tabular binary data.

\begin{algorithm}
\caption{Tabular Adversarial Training for binary classification with binary features}
\begin{algorithmic}[1]
\label{alg:tab-adv-train}
\REQUIRE Optimization algorithm $\mathcal{O}$, training samples $X$, number of epochs $N_{ep}$, index feature set $\Gamma \subset \mathbb{N}$, batch replay steps $m$, number of features to select $k$, feature selection criteria $s_{\Gamma,k}: \mathbb{R}^d \rightarrow \{0,1\}^d$
\STATE Initialize $\theta$
\STATE $\delta \leftarrow 0$ \COMMENT{\textit{$\delta \in \{-1,0,1\}^d$}}
\FOR{epoch = 1 to $N_{ep}/m$}
    \FOR{minibatch $B \subset X$}
        \FOR{$i = 1$ to $m$}
            \STATE \textit{Update $\theta$ using gradient-based optimizer $\mathcal{O}$}
            \STATE $g_\theta \leftarrow \mathbb{E}_{(x,y) \in B}[\nabla_\theta \ l(\text{clip}(x + \delta,0,1), y, \theta)]$
            \STATE $\theta \leftarrow \mathcal{O}(g_\theta)$
            \STATE \textit{Update $\delta$ with computed gradients $g_{adv}$}
            \STATE $g_{adv} \leftarrow \nabla_x \ l(\text{clip}(x + \delta,0,1), y, \theta)$
            \STATE $\gamma \leftarrow s_{\Gamma,k}(\Gamma, x)$ \COMMENT{\textit{$\gamma \in \{0,1\}^d$}}
            \STATE $\delta \leftarrow \delta + [\text{sign}(g_{adv})] \odot \gamma ]$ \COMMENT{\textit{Update only selected features}}\\
            \IF{$\sum_{i} abs(\delta_i)>k$}
            \STATE randomly  set $\sum_{i} abs(\delta_i) - k$  indices  to $0$  \COMMENT{\textit{Ensure modified features remain $\leq k$}}
            \ENDIF
            \STATE $\delta \leftarrow \text{clip}(\delta, -1, 1)$
        \ENDFOR
    \ENDFOR
\ENDFOR
\end{algorithmic}
\end{algorithm}

Algorithm \ref{alg:tab-adv-train} presents the adapted version of Shafahi et al's adversarial training method \cite{10.5555/3454287.3454589} for tabular data with binary features, i.e., where each input feature takes values in $\{0,1\}$. The model parameters, $\theta$,  are initialized according to the standard initialization scheme associated with a predefined optimization algorithm $\mathcal{O}$ \cite{ruder_overview_2016}. This could be random initialization, or more advanced alternatives \cite{narkhede_review_2022} for artificial neural network parameters, or setting parameters to specific values. In line 2, the adversarial perturbation, $\delta \in \mathbb{R}^d$, is initialized to a zero vector. This vector will accumulate the discrete perturbations applied to the input feature. Note that an alternative to this, could be to reset it each time a batch replay is completed. The algorithm iterates a total of $N_{ep}/m$ epochs (line 3). $N_{ep}/m$ represents the total desired number of training epochs, and $m$ is the number of batch replay steps performed within each batch.

For each epoch, the training data $X = \{x_1, x_2,...,x_N\}$ is divided into mini-batches $B \subset X$ (line 4). The algorithm iterates through each mini-batch $B$, and within each, it performs $m$ steps of adversarial training. This inner loop refines iteratively the adversarial perturbation at the same time it updates model parameters (line 4 to 17). Note that the adversarially perturbed samples, $x + \delta$, are constrained to have values within $[0,1]$, due to the nature of tabular binary data. This can be relaxed by not clipping $x + \delta$ (line 7).

In lines 11 to 12, a feature selection criterion is applied to the index feature set $\Gamma$. In the context of Android applications and DREBIN features, this eligibility is constrained by the semantics of the features: some features cannot be safely removed without breaking the app's functionality, but they can be added without adverse effects. For instance, adding unused permissions or declaring additional components in the manifest does not impact the APK's behavior whereas removing essential permissions or components could render the APK non-functional. Therefore, the adversarial perturbations are restricted accordingly. 

The selection function, $s_{\Gamma,k}$, takes the set of feature indices, $\Gamma$, and the input sample, $x$, and outputs a binary mask, $\gamma \in \{0,1\}^d$. Note that the mask has the same dimensionality as the input. Which features are selected in each step is implicitly controlled by the design of $s_{\Gamma,k}$. The criterion $s_{\Gamma,k}$ could be based on various heuristics. In the analysis conducted in Section \ref{sec:experimental-analysis}, we experiment with  $k$ random ($random$) selection and top $k$ ($topk$) selection based on the absolute value of gradients, $g_{adv}$. We also experiment with the maximum number of features to modify, $k$. 

In line 12, the algorithm computes the dot product between $sign(g_{adv}) \in \{-1,0,1\}^d$ and $\gamma \in \{0,1\}^d$, which preserves the value in the selected features and sets to zero the rest of indices in $sign(g_{adv})$. The perturbation, $\delta$, is then updated with this term. In line 13, we correct the difference between the number of perturbed indices and the upper bound, $k$, by randomly setting indices back to zero (no perturbation for that feature).

Lastly, in line 16 we clip the perturbation after adding the update for selected features in line 12. By doing so, we restrict $\delta$ to have component values in the set $\{-1,0,1\}$. This avoids having large absolute values in $\delta$ that would harden changing the direction of the perturbations in subsequent training steps.

\subsubsection{Label Smoothing} \label{sec:distillation}
Discrete true labels enforce strict class separation, resulting in sharp decision boundaries that are brittle under conditions like sparse features and scarce training data, which strives the well-studied curse of dimensionality \cite{goos_surprising_2001}. Under these conditions, the effectiveness of adversarial attacks is increased \cite{PAKNEZHAD2022178}.
To mitigate this, we adopt label smoothing, a regularization technique that replaces discrete labels with fuzzy labels that has been shown to improve the generalization of neural networks  \cite{10.5555/3454287.3454709}. In particular, we convert the binary classification task to a prediction task, where we want to predict the ``maliciousness" scores instead of true labels. To this end, we implement a smoothing labeling mechanism governed by hyperparameter $\lambda \in [0,1]$, that induces fuzziness into the true labels by making use of the probability predictions, ${f_s}(x)\in [0,1]$, of another classifier, $f_s$. Mathematically, the resulting label, $y_s \in [0,1]$, can be formulated as:

\begin{equation} \label{eq:label-smoothing}
    y_s : = (1-\lambda) y + \lambda f_s(x),
\end{equation}

where $x$ is an input sample and $y\in \{0,1\}$ is the discrete true label. This is similar to carrying out knowledge distillation \cite{yuan_revisiting_2020}, but in this case using a simpler learner and only partly distilling its knowledge captured during training. Instead of transferring knowledge from a complex teacher, we propose a simpler, explainable model, $f_s$, to introduce a degree of uncertainty or ``fuzziness" into the original discrete true label. The fusion hyperparameter, $\lambda$, controls the influence of $f_s(x)$ on the final smoothed target $y_s$.

\subsection{Winning Configuration in the IEEE SaTML'25 competition}
\label{sec:winning_configuration}
For the IEEE SaTML'25, we designed a particular configuration of the multi-step architecture introduced in Section \ref{sec:multi-step-theory}, that we presented under the name of \textbf{DeepTrust}. The candidate achieves a balance between robustness and efficiency by relying on only two distinct models with identical architectures but divergent training regimes. These models are arranged in a multi-step sequence of detectors, $F := \{f_1, f_2, f_3\}$, where:
\begin{itemize}
    \item $f_1$: \textbf{SAdvNet}, a deep neural network trained with a strong configuration of tabular adversarial training and label smoothing. This model is the most robust in the sequence. Adversarial training (cf. Algorithm \ref{alg:tab-adv-train}) is configured with $\{m=10$, $k=100, s_{\Gamma,k}=topk\}$. Label smoothing hyperparameter is set to $\lambda=0.5$ (Equation \ref{eq:label-smoothing}).      
    \item $f_2$: \textbf{wAdvNet}, a weakly adversarially trained network without label smoothing. While less robust than $f_1$, it is tuned to achieve a lower false positive detection rate, thus contributing complementary behavior. Adversarial training is configured with $\{m=2, k=75,s_{\Gamma,k}=topk\}$.   
    \item $f_3$: \textbf{SAdvNet}, reused in the third step of the sequence. Alternating the same robust model forces an adversary to readapt for a third time adversarial perturbations blindly. Model reuse raises the difficulty of successful evasion while adding negligible computational time as prediction is already precomputed in the first step.
\end{itemize}
The decision logic of the system is governed by a set of activation conditions, $C = \{\mathcal{C}_1,\mathcal{C}_2\}$:
\begin{equation}
    \begin{aligned}
        \mathcal{C}_1(x) &= [f_1(x) \geq 0.78], \\
        \mathcal{C}_2(x) &=  \mathcal{C}_{2,1}(x) \vee \mathcal{C}_{2,2}(x), \text{where} \\
        \mathcal{C}_{2,1}(x) &= [f_2(x) \geq 0.5] \text{ and} \\
        \mathcal{C}_{2,2}(x) &= [f_2(x) < 0.5 \ \wedge \ a(x) \geq 0.5], \\
    \end{aligned}
    \label{eq:activation-conditions}
\end{equation}
where $a(x)$ denotes an Isolation Forest \cite{liu_isolation-based_2012} anomaly detector trained with benign sample embeddings produced by the last hidden-layer from $f_2$. This auxiliary detector allows $\mathcal{C}_2$ to mitigate false positives while recovering true detections in cases of low classifier confidence.

The full system is defined as $\mathcal{F} = (F, C, t=0.5)$. The sequential decision process works as follows for a given input $x\in\mathbb{R}^d$:  
\begin{enumerate}
    \item If $\mathcal{C}_1(x)$ is satisfied, the system outputs $f_1(x)$.  
    \item Otherwise, $\mathcal{C}_2(x)$ is evaluated. $\mathcal{C}_2(x)$ is satisfied if either $\mathcal{C}_{2,1}(x)$ or $\mathcal{C}_{2,2}(x)$ are. If so, the output is given by $f_2(x)$, with anomaly detection serving as a fallback in cases of low classifier confidence.  
    \item If neither $\mathcal{C}_1$ nor $\mathcal{C}_2$ holds, the system defaults to $f_3$, which is identical to $f_1$. 
    \item At decision time, if $\mathcal{F}(x) < 0.5$, $x$ is classified as benign. Otherwise, $x$ is classified as malicious.
\end{enumerate}
This design ensures robustness against evasion attacks through the use of multiple strong models while maintaining a false positive rate below $1\%$ on non-perturbed goodware samples, in accordance with competition requirements.

The construction of DeepTrust followed a staged methodology. We first optimized a base MLP architecture, then searched for optimal adversarial training hyperparameters to generate a pool of robust models, and finally introduced label smoothing variations. An incremental search process guided by Bayesian optimization was applied at each stage and a Random Forest was optimized to generate smoothed labels. The final multi-step configuration was selected manually from this pool, combining strong and weak variants with carefully tuned thresholds and Isolation Forest contamination hyperparameter. An expanded description of the methodology employed to derive this configuration is given in Section \ref{sec:winning-configuration}.

\section{Evaluation}
\label{sec:evaluation}

This section details the evaluation methodology and results, and is organized as follows: first, we introduce the European Lighthouse on Secure and Safe AI's Robust Android Malware Detection (ELSA-RAMD) benchmark (Section \ref{sec:elsa-cyber}). Second, we conduct an empirical analysis using Track 1 to study the contribution to robustness of each component within our DeepTrust framework, comparing its performance to that of standard ensemble and individual models (Section \ref{sec:experimental-analysis}). Third, we present the final results from the IEEE SaTML'25 competition (Section \ref{sec:competition-results}).

\subsection{Robust Android Malware Detection Benchmark}\label{sec:elsa-cyber}
The Robust Android Malware Detection Benchmark is a competition organized by the European Lighthouse on Secure and Safe AI (ELSA) at the IEEE Conference SaTML 2025. The competition's goal is to assess the efficacy of ML-based methods for Android malware detection under feature-space, problem-space attacks and data drift based on the rules in Table \ref{tab:competition_rules}. 
\begin{table}[ht!]
    \centering
    
    \caption{Overview of ELSA-RAMD competition rules, including primary and tie-breaking metrics, and prerequisites for participation.     
    FSA: feature-space attack - PSA: problem space attack - TPR: true positive rate - FPR: false positive rate - AUT: area under time \cite{pendlebury_tesseract_2019}}
    
    \begin{tabular}{p{0.10\columnwidth} p{0.20\columnwidth} p{0.20\columnwidth} p{0.10\columnwidth}}  
    \toprule
    \textbf{Track} & \textbf{Winning Metric} & \textbf{Tie-Breaking} & \textbf{Requisite} \\
    \midrule
    Track 1 & TPR 100-FSA & (1) TPR 50-FSA; \newline (2) TPR 25-FSA; \newline (3) TPR no attack; \newline (4) FPR & FPR$\leq 1\%$ \\
    \midrule
    Track 2 & TPR 100-PSA & (1) TPR no attack;
    
    (2) FPR & FPR$\leq 1\%$ \\
    \midrule
    Track 3 & AUT on F1 Scores & - & - \\
    \bottomrule
    \end{tabular}
    \label{tab:competition_rules}
\end{table}

Accordingly, the competition is organized into the three following tracks:

\subsubsection{Track 1 - Adversarial Robustness to Feature-space
Attacks}
This track evaluates the robustness of the detection system against adversarial manipulations applied directly to feature vectors extracted following the DREBIN feature extraction methodology \cite{DBLP:conf/ndss/ArpSHGR14}, yielding 1,461,078 sparse binary features or a subset thereof. The attacker's model is as described in Section \ref{sec:threat-model}. Submissions to Track 1 must achieve a False Positive Rate (FPR) $\leq 1\%$ on a benign-only held-out test set before robustness is assessed (not meeting this prerequisite excludes candidates from robustness evaluation). The performance of models that satisfy this criterion is then measured by the True Positive Rate (TPR) under perturbation budgets of 0, 25, 50, and 100, with the winner determined by the robustness at 100 feature modifications. We denote each perturbation strength scenario as \{25,50,100\}-FSA. For more details on the feature space attack, we refer the reader to ~\ref{sec:feature_space_attack}.  

\subsubsection{Track 2 - Adversarial Robustness to Problem-space Attacks}
In this track, the attacker does not have knowledge of the model or its features, and it directly manipulates APK files through functionality-preserving modifications. This track allows defenders to perform their own feature extraction from APKs (i.e., are not restricted to the DREBIN features anymore) and, as in Track 1, must first satisfy the FPR $\leq 1\%$ constraint on the benign test set.  Robustness is then measured by TPR on adversarial malware, with at most 100 features adversarially modified, denoted as 100-PSA. Further details of the problem-space attack are covered in ~\ref{sec:problem_space_attack}. 

\subsubsection{Track 3 - Temporal Robustness to Data Drift}
In Track 3, it is measured the performance decay over time of malware detectors due to the natural evolution of goodware and malware. Models must accept APK files as input and are evaluated on four temporally distinct test sets spanning 2020–2022. The primary metric is the Area Under Time (AUT) \cite{pendlebury_tesseract_2019}, which summarizes the F1 scores obtained across successive evaluation rounds. Unlike Tracks 1 and 2, there is no explicit FPR constraint and the winner is the model with the highest AUT.  

\begin{table}[ht!]
    \centering
    \caption{Overview of the datasets used for training and testing across all benchmark tracks.}
    \begin{tabular}{p{0.20\columnwidth} p{0.10\columnwidth} p{0.10\columnwidth} p{0.05\columnwidth}}
    \toprule
    \textbf{Dataset} & \textbf{Timeframe Sampled} & \textbf{Size} & \textbf{Ratio G:M} \\
    \midrule
    Training Set & 2017–2019 & 75K & 9:1 \\
    \midrule
    \multicolumn{4}{l}{\textit{Track 1, 2}} \\
    Test Set - Goodware & 2020–2022 & 5K & - \\
    Test Set - Malware & 2020–2022 & 1.25K & - \\
    \midrule
    \multicolumn{4}{l}{\textit{Track 3}}\\
    Test Set - Round 1 & 2020 & 25K & 9:1 \\
    Test Set - Round 2 & 2020–2021 & 25K & 9:1 \\
    Test Set - Round 3 & 2021 & 25K & 9:1 \\
    Test Set - Round 4 & 2021–2022 & 25K & 9:1 \\
    \bottomrule
    \end{tabular}
    \label{tab:dataset_overview}
\end{table}

\subsubsection{Datasets and Evaluation Metrics.}
The characteristics of the datasets for training and testing are listed in Table \ref{tab:dataset_overview}.
For Tracks 1 and 2, performance is assessed using the True Negative Rate (TNR) and the True Positive Rate (TPR).
The True Negative Rate, also known as specificity, measures the model's ability to correctly identify benign applications, thereby minimizing the false positives. TNR is defined as:
\[
TNR = \frac{TN}{TN + FP},
\]
where $TN$ (True Negatives) are the benign apps correctly classified as benign, and $FP$ (False Positives) are the benign apps incorrectly classified as malware. The True Positive Rate (TPR), also known as recall or detection rate,
measures the model's ability to successfully detect malicious applications. TPR is defined as:
\[
TPR = \frac{TP}{TP + FN},
\]
where $TP$ (True Positives) are the malware samples correctly identified as malware, and $FN$ (False Negatives) are malware samples that the model fails to detect.

In Track 1, TPR is evaluated under no attack and against 25-FSA, 50-FSA, 100-FSA. In Track 2, TPR is evaluated under no attack and against 100-PSA. Track 3 evaluates the model's resilience against \emph{concept drift} (evolution of both malware and goodware over time). The primary evaluation metric is AUT \cite{pendlebury_tesseract_2019}, which summarizes model performance across the four temporally distinct test rounds Test Set Track 3 - Round 1, 2, 3, 4 (Table \ref{tab:dataset_overview}).

\subsection{Ablation Study} \label{sec:experimental-analysis}
Next we conduct an experimental assessment of the contribution to robustness of each component in DeepTrust. Our findings show that (1) tabular adversarial training (Section \ref{sec:tab-adv}) and label smoothing (Section \ref{sec:distillation}) improve robustness to evasion attacks with no performance degradation; (2) MLPs trained through adversarial training and label smoothing converge to parameters that encode substantially different dense representations with respect to a MLP with equal architecture and non-robust training (vanilla-MLP); (3) greater diversity of the learned data representation among internal models increases the robustness of the multi-step strategy; and (4) multi-step strategies with diverse models outperform both individual robust models and ensemble-based detectors.

\textbf{Experimental Setup.} 
We evaluate the components of DeepTrust under the ELSA-RAMD Track 1 framework, which simulates a worst-case evasion scenario where the attacker has knowledge of the features and the output probabilities of the detection system, to estimate the upper bounds of the system's degradation. Vanilla-MLP  is a feed-forward multilayer perceptron, compound of two hidden layers with 128 and 64 units, leaky ReLU \cite{xu_empirical_2015} activation function in the hidden units, and sigmoid function for the output unit. Due to the imbalance ratio between goodware and malware (9:1) in the training set, the binary cross-entropy loss is weighted. The model is trained using Adam with learning rate set to $0.001$, $\beta_1 = 0.99$, $\beta_2 = 0.999$ and $\epsilon = 1 \cdot 10^{-8}$ \cite{kingma_adam_2014} for 10 epochs, batch size 32, on 80\% of the training partition, all starting with identical weights, whereas the remaining 20\% is used as validation set to select best weights based on F1 Score. The 1\% FPR constraint is relaxed to study the robustness variability across components.

\subsubsection{Adversarial Training on Tabular Data}
\label{sec:experiments_adversarial_training}

We explore the effects of  the following hyperparameters on the performance of adversarial training:

\begin{itemize}
    \item Batch replay, $m$. It defines the number of times a batch is reused for crafting the iterative adversarial perturbations.
    \item Maxim feature changes, $k$. It defines the upper limit of altered features during training.
    \item Feature selection strategy, $s_{\Gamma,k}$. It defines the strategy used to select which features will be perturbed. We have defined two strategies: (1) $random$ selection and (2) $topk$ selection.
\end{itemize}

Adversarially trained models (adversarial-MLPs) implement the same architecture, initialization, and training configuration as the vanilla-MLP. Therefore, differences in their learned parameters and performance only come from the application of the tabular adversarial algorithm (Algorithm \ref{alg:tab-adv-train}).

\begin{figure}[ht]
    \centering
    \subfloat[TNR Test Set Tracks 1, 2 - Goodware]{
        \includegraphics[width=0.30\textwidth]{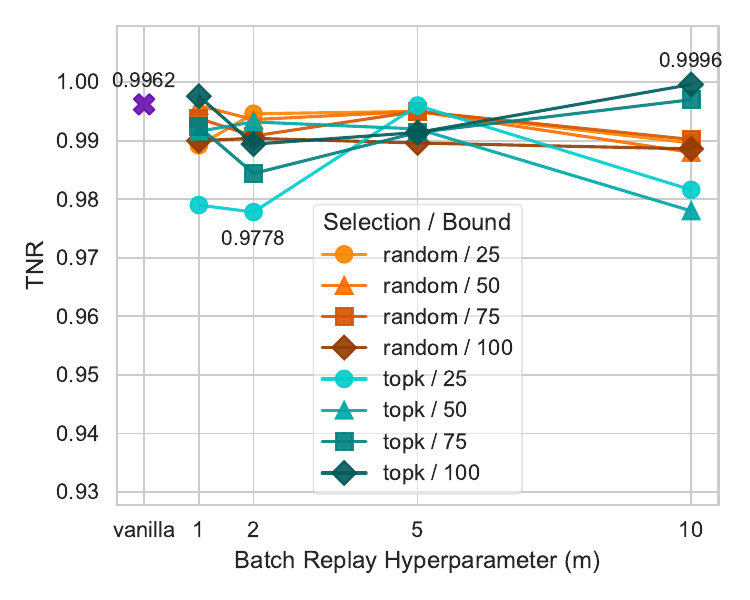}
        \label{fig:adv-train-a}
    }
    \hspace{1em}
    \subfloat[TPR Test Set Tracks 1, 2 - Malware (TSM)]{
        \includegraphics[width=0.30\textwidth]{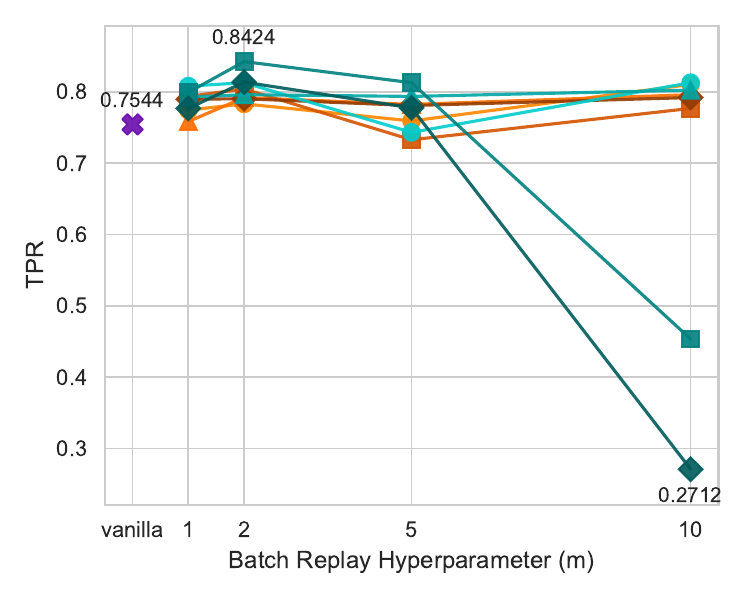}
        \label{fig:adv-train-b}
    }
    
    \subfloat[TPR on TSM under 25-FSA]{
        \includegraphics[width=0.30\textwidth]{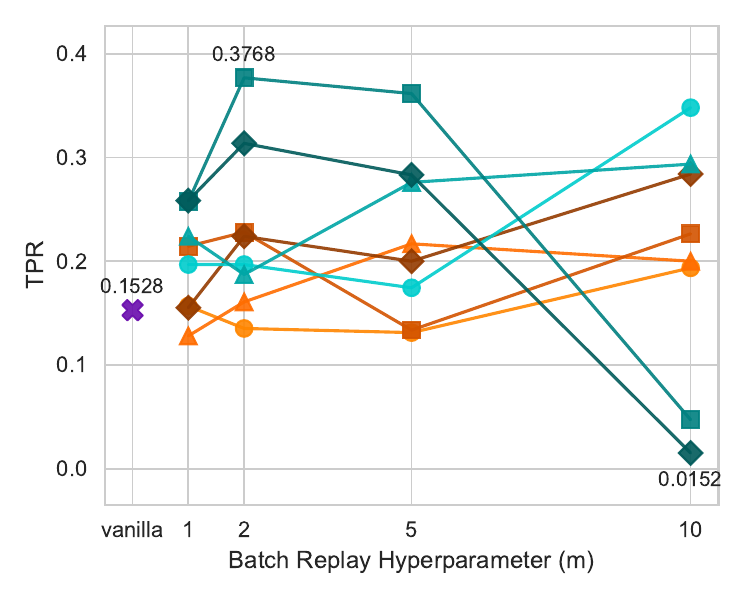}
        \label{fig:adv-train-c}
    }
    \hspace{1em}
    \subfloat[TPR TSM under 50-FSA]{
        \includegraphics[width=0.30\textwidth]{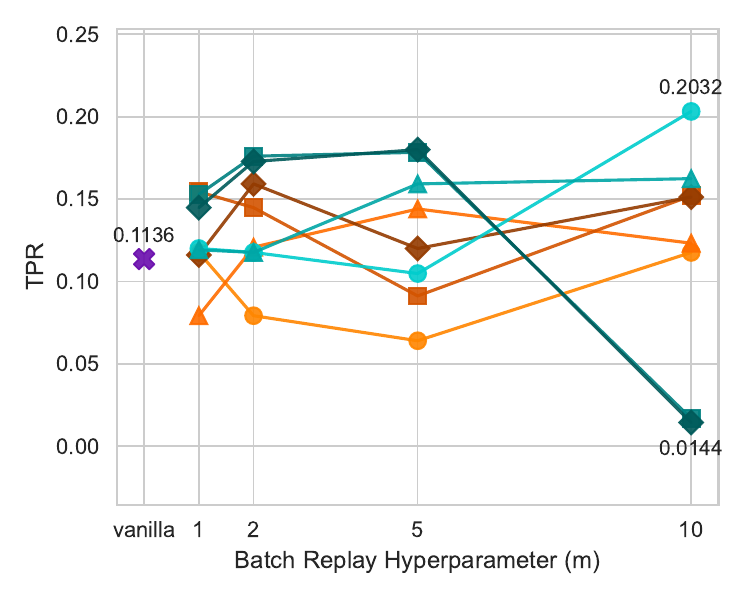}
        \label{fig:adv-train-d}
    }

    \subfloat[TPR TSM under 100-FSA]{
        \includegraphics[width=0.30\textwidth]{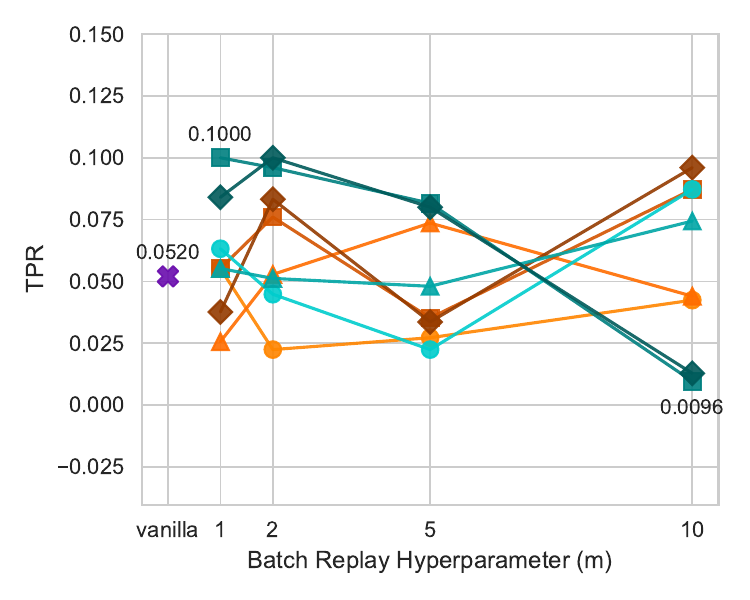}
        \label{fig:adv-train-e}
    }

    \caption{Tabular adversarial training in ELSA-RAMD Track 1 (y-axis: recall/specificity, x-axis: $m$).}
    \label{fig:tab-adv-train}
\end{figure}

Figure \ref{fig:tab-adv-train} show the evolution of TNR and TPR with increasing batch replay, $m$, for different $s_{\Gamma,k}$ and $k$ combinations. Under no attack (Figure \ref{fig:tab-adv-train}-A, B), most configurations achieve similar TNR and TPR as the vanilla-MLP, showing that adversarial training does not harm the performance of the models on clean data. However, for larger $k$ values ($k \in \{75,100\}$) and high batch replay ($m=10$), TPR drops below 0.5 (Figure \ref{fig:tab-adv-train}-B), likely because the overly aggressive perturbations create unrealistic, out-of-distribution examples that hinder learning convergence.

Against FSA (Figures \ref{fig:tab-adv-train}C-E), $random$ feature selection, ($s_{\Gamma,k}=random$), benefits from high batch replay (e.g., $m=10$) and large feature-change bounds ($k\in\{75,100\}$), outperforming vanilla-MLP by a large margin, e.g 186\% under 25-FSA, 133\% under 50-FSA and 185\% under 100-FSA in the case of adversarial-MLP with configuration set $\{m=10,k=100,s_{\Gamma,k}=random\}$. This is because its stochastic nature requires stronger setups to craft effective adversarial examples during training, as generate the adversarial perturbations by modifying the most sensitive features. By contrast, $topk$ feature selection, which ranks features by the absolute value of the loss gradients, yields the highest robustness. Since this selection strategy explicitly targets the most sensitive features, very high batch replay, $m$, can destabilize learning when combined with large modification bounds, $k$, while smaller $k$ values benefit from higher $m$. Therefore, across all evasion attack scenarios, i.e., \{25,50,100\}-FSA, these two optimal $topk$ configurations (low $m$ with high $k$, or high $m$ with low $k$) outperform vanilla-MLP, with configurations yielding TPR gains of 247\% under 25-FSA for adversarial-MLP with $\{m=2,k=75,s_{\Gamma,k}=topk\}$, 179\% under 50-FSA for $\{m=10,k=25,s_{\Gamma,k}=topk\}$ and 192\% under 100-FSA for $\{m=1,k=75,s_{\Gamma,k}=topk\}$.
Note that these gains are achieved while maintaining TNR above 0.98 (Figure \ref{fig:tab-adv-train}-A).

\subsubsection{Label Smoothing}
\label{sec:experiments_label_smoothing}
\begin{figure}[ht]
    \centering
    \subfloat[TNR on Test Set Tracks 1 \& 2 - Goodware; TPR on Test Set Tracks 1 \& 2 - Malware (TSM)]{
        \includegraphics[width=0.30\textwidth]{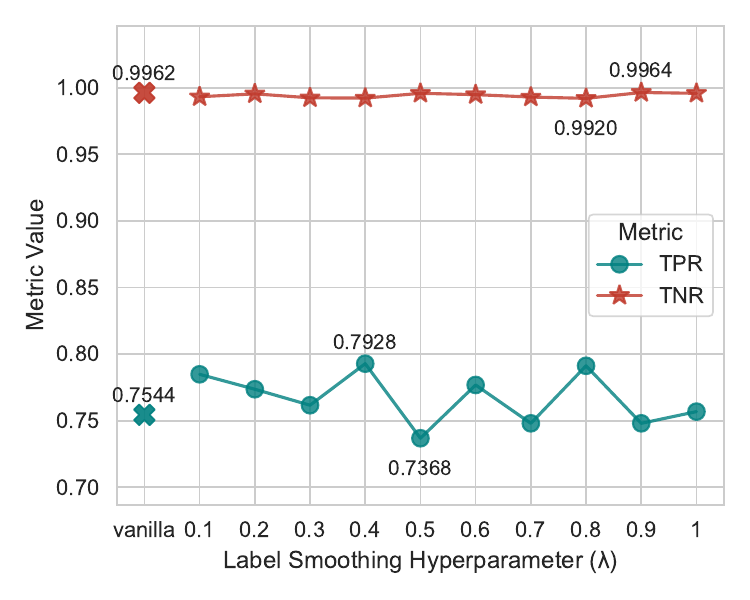}
        \label{fig:ls-a}
    }
    \hspace{2em}
    \subfloat[TPR on TSM under \{25,50,100\}-FSA.]{
        \includegraphics[width=0.30\textwidth]{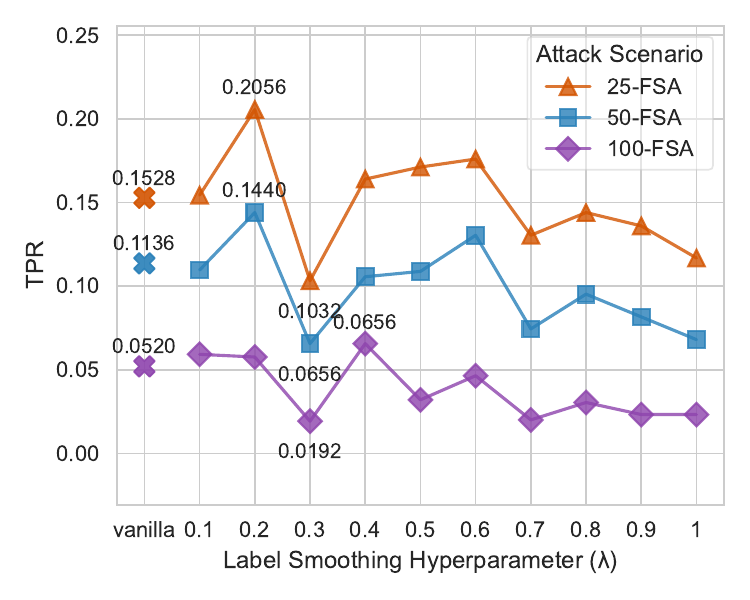}
        \label{fig:ls-b}
    }

    \caption{Label Smoothing in ELSA-RAMD Track 1 (y-axis: TPR/TNR, x-axis: $\lambda$).}
    \label{fig:label-smoothing}
\end{figure}

We examine the effect of varying the label smoothing hyperparameter, $\lambda$, using a Random Forest as $f_s$ in Equation \ref{eq:label-smoothing}. This detector has been tuned by searching for optimal hyperparameters following the methodology described in Stage 3, Section \ref{sec:competition-results}. Label smoothed models (smoothed-MLPs) implement the same architecture, initialization, and training configuration as vanilla-MLP. Therefore, differences in their learned parameters and performance only come from the application of label smoothing, to different degrees, to true labels during training. Results show that the  performance of smoothed-MLPs under no attack remains close to the vanilla-MLP across $\lambda$ values (Figure \ref{fig:label-smoothing}-A). However, high $\lambda$ values reduce their robustness (Figure \ref{fig:label-smoothing}-B), as the smoothed labels rely more on $f_s$ predictions than on true informative labels of the samples. Nevertheless, small $\lambda$ values (e.g., $\lambda \in \{0.1, 0.2\}$) improve robustness against FSA attacks.

\subsubsection{Multi-Step Strategy}\label{sec:multi-step}
The multi-step classification framework, presented in Section \ref{sec:deeptrust}, is based on the assumption that using a sequence of models that project the original feature space to diverse dense representations improves robustness against adversarial attacks. Intuitively, this design forces the adversary to, sequentially and incrementally, deceive multiple models, each of which performs classification in a distinct latent projection. Before testing this assumption, we investigate whether the learned embedding space differs from that of the vanilla-MLP. Specifically, it is analyzed the global structure of the embeddings produced by the last hidden layer of the best models trained with tabular adversarial training (cf. Section \ref{sec:experiments_adversarial_training}) and label smoothing (cf. Section \ref{sec:experiments_label_smoothing}), alongside vanilla-MLP. Specifically, we select the following four models: adversarial(75)-MLP with parameter set $\{{m = 2, k = 75, s_{\Gamma,k} = topk}\}$, adversarial(100)-MLP with set $\{{m = 2, k = 100, s_{\Gamma,k} = topk}\}$, smoothed(0.2)-MLP with $\lambda=0.2$ and smoothed(0.4)-MLP with $\lambda=0.4$.

\begin{figure*}[ht!]
    \centering

    \newcommand{\gridwidth}{0.18\textwidth}

    \subfloat[vanilla-MLP - UMAP]{
        \includegraphics[width=\gridwidth]{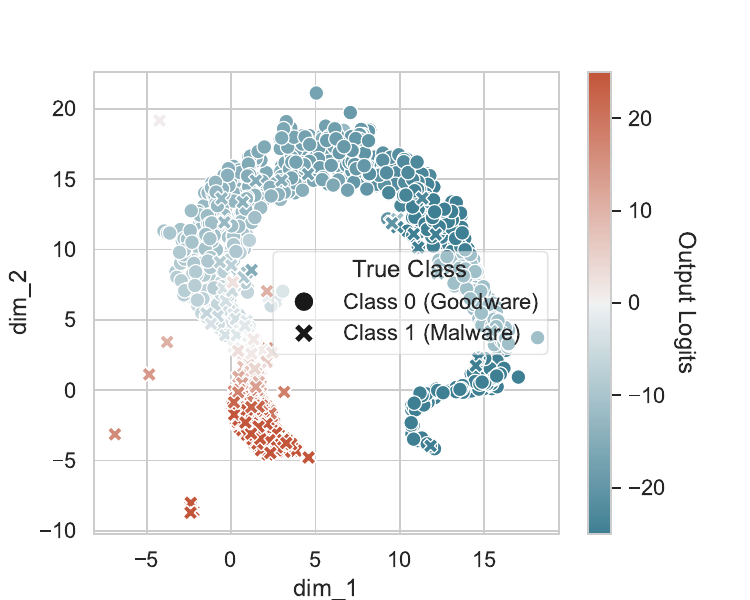}
        \label{fig:embed-a}
    }\hfill
    \subfloat[adversarial(75)-MLP - UMAP]{
        \includegraphics[width=\gridwidth]{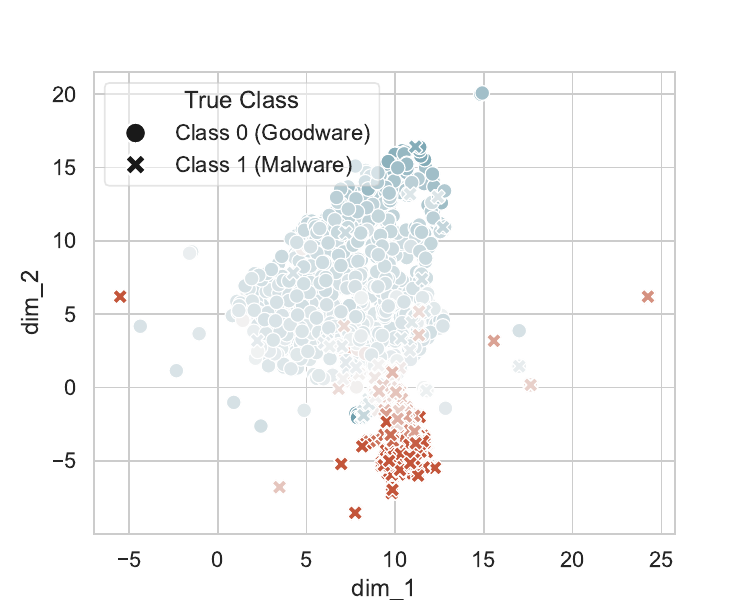}
        \label{fig:embed-c}
    }\hfill
    \subfloat[adversarial(100)-MLP - UMAP]{
        \includegraphics[width=\gridwidth]{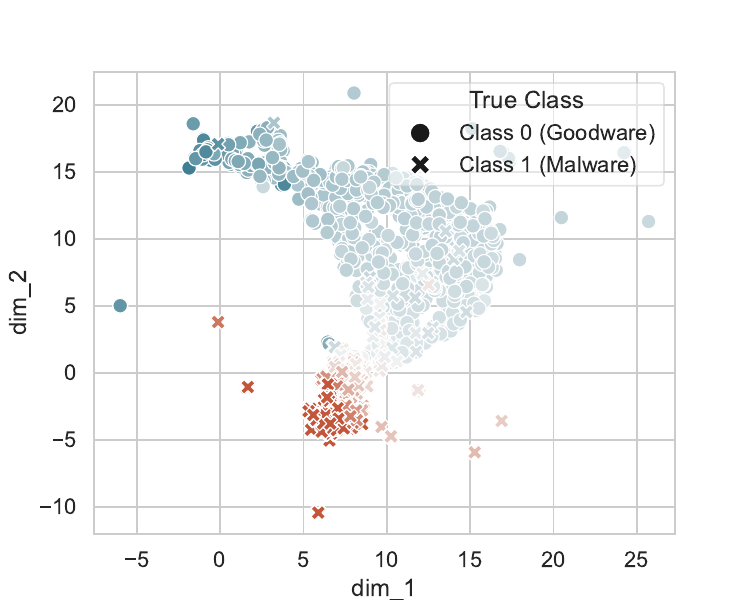}
        \label{fig:embed-e}
    }\hfill
    \subfloat[smoothed(0.2)-MLP - UMAP]{
        \includegraphics[width=\gridwidth]{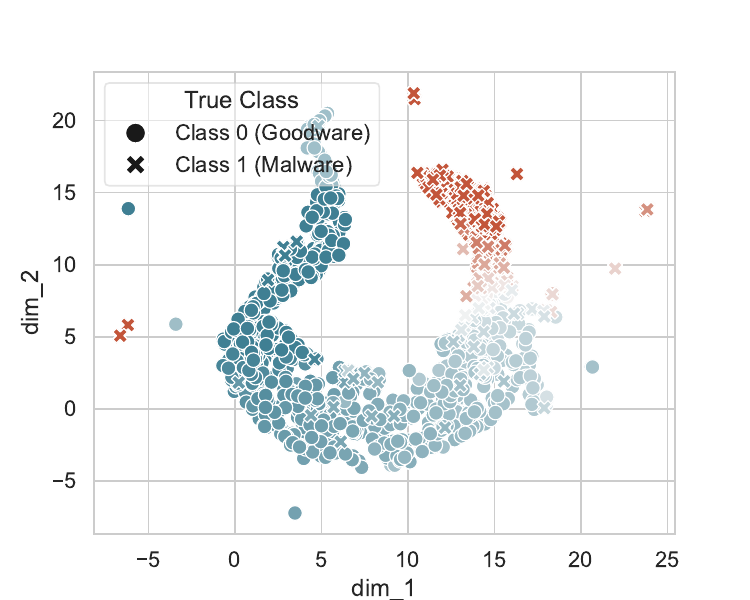}
        \label{fig:embed-g}
    }\hfill
    \subfloat[smoothed(0.4)-MLP - UMAP]{
        \includegraphics[width=\gridwidth]{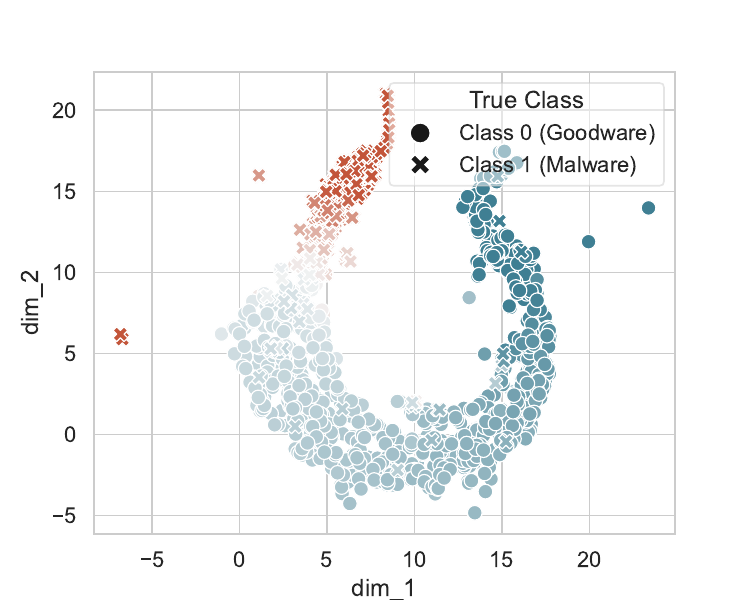}
        \label{fig:embed-i}
    }
    \\[\medskipamount]

    \subfloat[vanilla-MLP - t-SNE]{
        \includegraphics[width=\gridwidth]{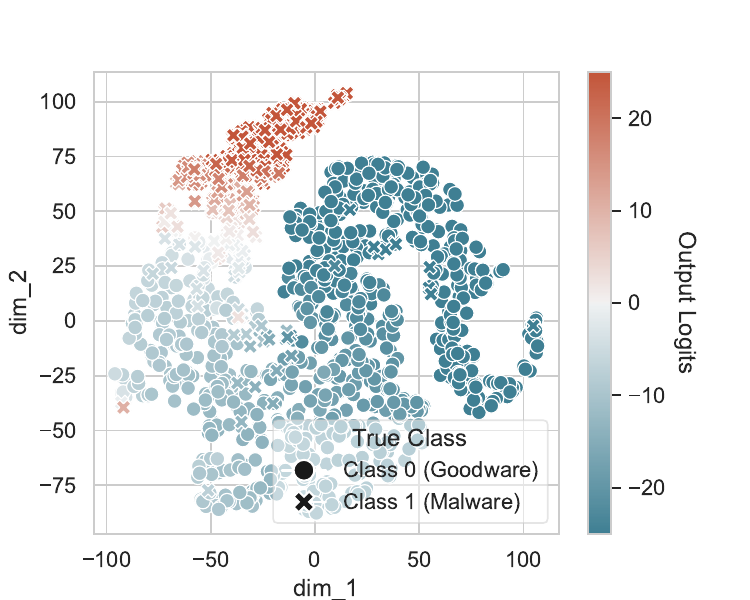}
        \label{fig:embed-b}
    }\hfill
    \subfloat[adversarial(75)-MLP - t-SNE]{
        \includegraphics[width=\gridwidth]{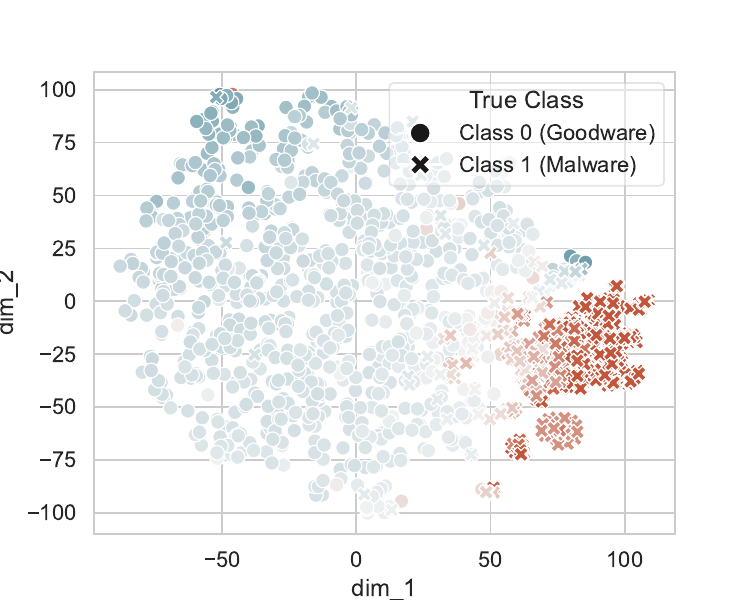}
        \label{fig:embed-d}
    }\hfill
    \subfloat[adversarial(100)-MLP - t-SNE]{
        \includegraphics[width=\gridwidth]{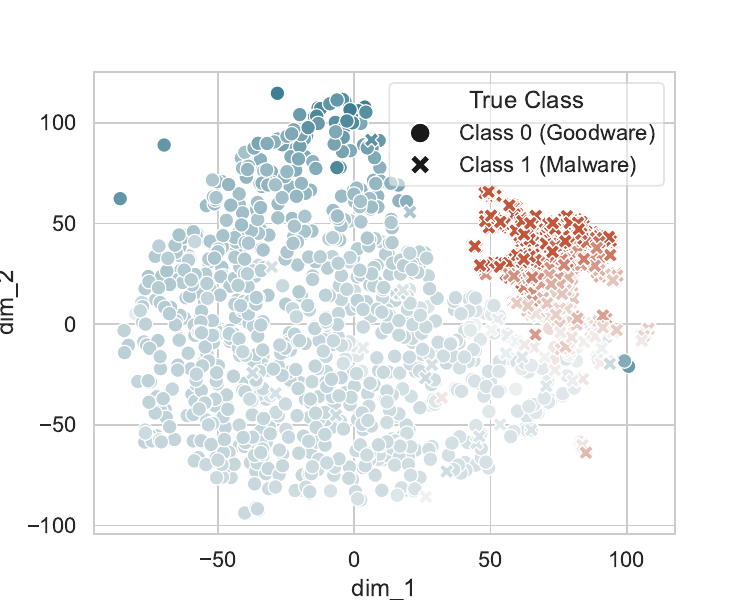}
        \label{fig:embed-f}
    }\hfill
    \subfloat[smoothed(0.2)-MLP - t-SNE]{
        \includegraphics[width=\gridwidth]{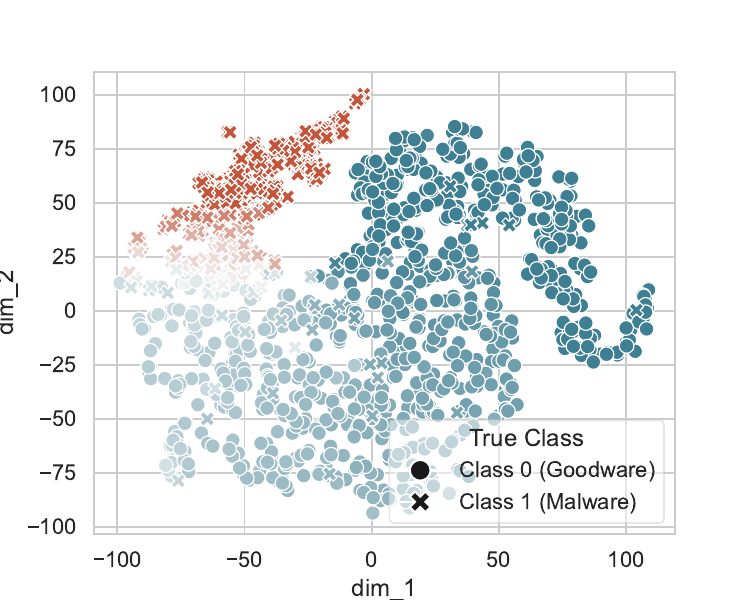}
        \label{fig:embed-h}
    }\hfill
    \subfloat[smoothed(0.4)-MLP - t-SNE]{
        \includegraphics[width=\gridwidth]{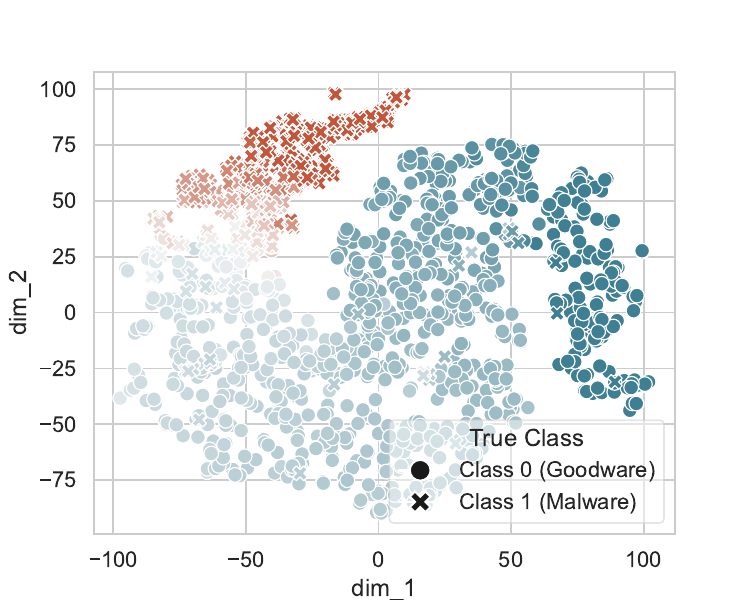}
        \label{fig:embed-j}
    }

    \caption{2D visualization of embeddings. The top row shows UMAP projections and the bottom row shows t-SNE projections for five different models (columns). The color gradient red-white-blue represents logit prediction values.}
    \label{fig:embedding-projections}
\end{figure*}

\textbf{Visual Representation.}
We apply t-SNE \cite{JMLR:v9:vandermaaten08a} and UMAP \cite{McInnes2018} to the validation set data embeddings, denoted also as $x \in \mathbb{R}^{64}$ for notation simplicity. As shown in Figure \ref{fig:embedding-projections}, the compressed projection of the embeddings from models trained with similar schemes tend to form a space with a resembling global structure, e.g., smoothed(0.2)-MLP and smoothed(0.4)-MLP, or adversarial(75)-MLP and adversarial(100)-MLP. Furthermore, latent feature vectors from smoothed-MLPs exhibit a distribution more closely resembling that of the vanilla-MLP than adversarial-MLPs.  

\textbf{Prototypes.} Instead of visualizing the latent features, we can interpret the weight vector of the classification layer (output layer with 1 unit) as a learned template or prototype of positiveness (maliciousness in our case) \cite{muller_when_2019}. This can be drawn from the fact that the logits, obtained from the dot product between the output weight vector and the embedding produced by the last hidden layer, $x^Tw$, appear as a subtraction in the second term of the expression of the distance between the weight vector and the embedding itself: 
$$
\|x - w\|^2 = x^T x - 2x^T w + w^T w.
$$

Thus, the closer an example's representation is to the template, the smaller its Euclidean distance will  be. Since this distance is always positive, any decrease corresponds to an increase in the logits. We therefore examine correlations between example logits (Table \ref{tab:model-comparison-results}). Since all models are trained for the same binary malware detection task (i.e., same data and features), the correlation of their logits with respect to the vanilla-MLP is reasonably high. Nonetheless, the logits of the models trained with label smoothing show greater correlation to the vanilla-MLP than adversarial-MLPs. This is aligned with the 2D visualization of the latent features, where those extracted from the vanilla-MLP and the smoothed-MLPs show a much more similar global structure compared to those from adversarial-MLPs (Figure \ref{fig:embedding-projections}).

\begin{table*}[ht!]
    \centering
    \caption{Evaluation in ELSA-RAMD Track 1. Global Best/second-best are bold/underlined. $\rho$: Pearson correlation of logits in validation partition with respect to vanilla-MLP. TNR: True negative rate; TPR: True Positive Rate; FSA \textit{X}: Feature-Space Attack under X feature manipulation bound}.
    \begin{tabular}{p{0.35\textwidth} c p{0.06\textwidth} p{0.06\textwidth} p{0.06\textwidth} p{0.06\textwidth} p{0.06\textwidth}}
    \toprule
    & \textbf{$\rho$} & \textbf{TNR} & \textbf{TPR no FSA} & \textbf{TPR 25-FSA} & \textbf{TPR 50-FSA} & \textbf{TPR 100-FSA} \\
    \midrule
    \textbf{Detector No.1: vanilla-MLP} & 1 & \textbf{.9962} & .7544 & .1528 & .1136 & .0520 \\
    \midrule
    \textbf{Detector No.2: smoothed(0.2)-MLP} & .9655 & .9954 & .7736 & .2056 & .1440 & .0576    \\
    \textbf{Multi-step} $F=$ \{$f_1$:(2), $f_2$:(1), $f_3$:(2)\} & - & .9944 & .7848 & .2344 & .1592 & .0664 \\
    \textbf{Ensemble} \{$f_1$:(2), $f_2$:(1)\} & - & \textbf{.9962} & .7712 & .1584 & .1056 & .0440 \\
    \midrule
    \textbf{Detector No.3: smoothed(0.4)-MLP} & .9257 & .9922 & .7928 & .1640 & .1056 & .0656  \\
    \textbf{Multi-step} $F=$ \{$f_1$:(3), $f_2$:(1), $f_3$:(3)\} & - & .9912 & .8008 & .2216 & .1464 & .0720 \\
    \textbf{Ensemble} \{$f_1$:(3), $f_2$:(1)\} & - & \textbf{.9962} & .7808 & .1472 & .0952 & .0496 \\
    \midrule
    \textbf{Detector No.4: adversarial(75)-MLP} & .8157 & .9894 & .8136 & .3136 & .1728 & .1000 \\
    \textbf{Multi-step} $F=$ \{$f_1$:(4), $f_2$:(1), $f_3$:(4)\} & - & .9834 & \textbf{.8496} & \textbf{.4008} & \underline{.1952} & \textbf{.1056} \\
    \textbf{Ensemble} \{$f_1$:(4), $f_2$:(1)\} & - & .9948 & .7960 & .2128 & .1400 & .0576 \\
    \midrule
    \textbf{Detector No.5: adversarial(100)-MLP} & .8678 & .9844 & \underline{.8424} & \underline{.3768} & .1760 & .0960 \\
    \textbf{Multi-step} $F=$ \{$f_1$:(5), $f_2$:(1), $f_3$:(5)\} & - & .9880 & .8224 & .3712 & \textbf{.2112} & \underline{.1040} \\
    \textbf{Ensemble} \{$f_1$:(5), $f_2$:(1)\} & - & \underline{.9958} & .7752 & .2168 & .1384 & .0672 \\
    \bottomrule
    \end{tabular}
    \label{tab:model-comparison-results}
\end{table*}

\textbf{Results.} Here we empirically explore whether combining models with diverse internal representations improves the robustness against adversarial attacks. Table \ref{tab:model-comparison-results} reports the 
performance of different approaches using the selected detection models against \{25,50,100\}-FSA. For each of these, we create multi-step and ensemble variants. In all combinations, we use the non-robust vanilla-MLP as the fixed internal model. This allows us to measure, in isolation, the increase in robustness achieved solely through the diversity of the internal detectors. The multi-step approaches are defined as a cascade of classifiers, $F$, applied sequentially, as per Equation \ref{eq:multi-step}, where activation conditions, $\mathcal{C}_1$ and $\mathcal{C}_2$, are defined as follows:
\begin{equation}
    \begin{aligned}
        \mathcal{C}_1(x) &= \left[ f_1(x) \geq 0.75 \right]; \\
        \mathcal{C}_2(x) &= \left[ f_2(x) \geq 0.5 \right].
    \end{aligned}
\end{equation}

In contrast, ensemble approaches average the confidence scores of models $f_1$ and $f_2$ to reach a verdict. The class separation threshold, $t$, is set to $0.5$ in both cases.

The results in Table~\ref{tab:model-comparison-results} show that the multi-step strategy is consistently more robust under FSA compared to the corresponding individual models. In contrast, the ensemble-based approaches do not increase robustness but degrade performance, leading to a lower TPR under attack compared to the single detectors from which they are constructed. For instance, detector No. 3, smoothed(0.4)-MLP, achieves a TPR of $0.7928$ under no attack but the TPR drops sharply to $0.1640$ under 25-FSA. By contrast, multi-step with $F=$ \{$f_1$:(3), $f_2$:(1), $f_3$:(3)\} attains a TPR of $0.2216$ under attack, whereas the ensemble counterpart, \{$f_1$:(3), $f_2$:(1)\}, performs worse, yielding a TPR equal to $0.1472$, lower than both the single detector and the multi-step variants. This trend, where multi-step variants consistently demonstrate greater robustness than single and ensemble detectors, is most pronounced when model diversity is highest, a finding that corroborates our central hypothesis: sequentially stacking models that have learned dissimilar representations yields a more effective defense. The most compelling example is the multi-step with $F=$ \{$f_1$:(4), $f_2$:(1), $f_3$:(4)\}, which combines the adversarial(75)-MLP with vanilla-MLP. This configuration achieves the lowest logit correlation ($\rho=0.8157$) and the highest TPR under attacks (TPR equals to 0.4008, 0.1952 and 0.1056 for 25-FSA, 50-FSA, and 100-FSA, respectively). 

\subsection{IEEE SaTML'25 Competition Results} 
\label{sec:competition-results}
We now describe how we constructed the multi-step detection system that won the IEEE SaTML'25 competition and report its performance on all competition tracks \cite{ramd-elsa-competition}.

\subsubsection{Constructing the Winning Configuration of DeepTrust}
\label{sec:winning-configuration}
We followed a staged methodology to construct DeepTrust, with data splitting and standard training hyperparameters preset as per the experimental setup described in Section \ref{sec:experimental-analysis}:
\begin{itemize}
    \item \textbf{Stage 1 – Base architecture search:} We first optimized the hyperparameters of the MLP architecture with Optuna's Tree-Structured Parzen Estimator (TPE) \cite{akiba_optuna_2019} optimization algorithm, selecting the configuration that maximized the validation's F1 score over 20 trials with early stopping. The hyperparameter search explored the number of layers $\{2, 3\}$, the hidden layer sizes $\{32, 64, 128, 256\}$, ReLU/Leaky ReLU activations, dropout value in $[0.00, 0.75]$ with a step size of $0.05$, weight decay, and the positive class weight of the loss function. 
    \item \textbf{Stage 2 – Adversarial training search:} Using the optimal architecture found in Stage 1, we explore how the adversarial training hyperparameters impact the TNR and TPR against adversarial examples. In this stage, the optimization objective, $J$, is designed as:
        $$
J = \max\left(0, \frac{TNR - 0.95}{0.05}\right) \cdot \left( \prod_{\alpha} TPR_{\alpha\text{-FSA}} \right)^{\frac{1}{4}}.$$
    $J$ computes the geometric mean of the model's TPR on clean examples and adversarial examples generated with varying perturbation budgets, $\alpha \in \{0, 25, 50, 100\}$, on a randomly selected subset of the validation set, conditioned by a penalty function that drives the score to zero if the FPR on clean examples falls below 95\%. The hyperparameter search explored the number of batch replays, $m \in [2,20]$, the maximum number of features to be modified, $k$, in $[25,200]$ with a step size of $50$, and the feature selection function, $s_{\Gamma,k} \in \{topk,random\}$.
   
    \item \textbf{Stage 3 – Label smoothing search:} We optimized a Random Forest for $f_s$ in Equation \ref{eq:label-smoothing}. Similarly to Stage 1, TPE's optimization searched for the optimal hyperparameters for the Random Forest by maximizing its F1 score on the validation set over 20 trials. The hyperparameter search explored the number of tree estimators in $[25,100]$ with a step size of 5, the division criterion (Gini index/Shanon's entropy), the minimum number of samples required per leaf $[1,501]$ with a step size of 50, and the weight of the positive class. We then searched for $\lambda$ values in Equation \ref{eq:label-smoothing} that yielded robustness gains when combined with the adversarial training configuration found in Stage 2.
    
    \item \textbf{Stage 4 – Final assembly:} Lastly, we assessed a combination of models trained in previous stages to find the optimal combination, i.e., highest robustness against FSA and a false positive detection rate below or close to 1\% on clean examples. We then tuned manually the activation condition set $C=\{\mathcal{C}_1,\mathcal{C}_2\}$ (cf. Equation \ref{eq:activation-conditions}) and $t$. Concretely, combinations of different lower bounds for $\mathcal{C}_1$, the contamination parameter of the Isolation Forest model, $a$, in activation condition $\mathcal{C}_2$ and the class decision hyperparameter, $t$, to achieve under 1\% of TNR (prerequisite for the submission). The resulting configuration, summarized in Table \ref{tab:deeptrust-configuration-compact}, achieved 1st place in Track 1 of the IEEE SaTML'25 competition.
\end{itemize}

\begin{table*}[h!]
    \centering
    \caption{Final configuration of DeepTrust. Shared parameters are reported once.}
    \begin{tabular}{@{} l l c c @{}}
    \toprule
    \textbf{Category} & \textbf{Hyperparameter} & \textbf{Value (shared)} & \textbf{Diff. (SAdvNet / wAdvNet)} \\
    \midrule
    \textbf{Model Architecture} & Hidden Layers & (256, 32, 256) & - \\
     & Activation & Leaky ReLU & - \\
     & Dropout & 0.70 & - \\
    \midrule
    \textbf{Training} & Batch Size & 32 & - \\
     & Epochs & 10 & - \\
    \midrule
    \textbf{Optimizer (Adam)} & Learning Rate & 0.001 & - \\
     & Weight Decay & 0.00246 & - \\
    \midrule
    \textbf{Loss (BCE)} & Pos. Class Weight & 8.5 & - \\
    \midrule
    \textbf{Adversarial Training} & Batch Replays (m) & - & 10 / 2 \\
     & Number of features ($k$) & - & 100 / 75 \\
     & Feature Selection ($s_{\Gamma,k}$) & $topk$ & - \\
    \midrule
    \textbf{Label Smoothing} & $\lambda$ & - & 0.5 / 0.0 \\
    \midrule
    \textbf{Activation Conditions} & $C_1$ & \multicolumn{2}{c}{$f_1(x) \geq 0.78$} \\
     & $C_2$ & \multicolumn{2}{c}{$f_2(x) \geq 0.5 \ \vee \ (f_2(x) < 0.5 \wedge a(x)\geq 0.5)$} \\
     & $a(x)$ contamination & \multicolumn{2}{c}{0.14} \\
     & Threshold $t$ & \multicolumn{2}{c}{0.5} \\
    \bottomrule
    \end{tabular}
    \label{tab:deeptrust-configuration-compact}
\end{table*}

Tables \ref{tab:track-1-classification}, \ref{tab:track-2-classification} and \ref{tab:track-3-classification} report the results achieved by our sequential multi-step detection system on the IEEE SaTML'25 competition \cite{ramd-elsa-competition}. We have also included two additional ensemble-based detectors, namely XGBoost and Random Forest, to benchmark our multi-step detector against standard ensemble methods. In both cases, optimal configurations are found by following the methodology described in Stage 1.

\subsubsection{Track 1 - Adversarial Robustness to FSA}

The results of Track 1 (Table \ref{tab:track-1-classification}) clearly demonstrate DeepTrust's superior robustness against feature-space attacks. While maintaining the highest detection rate for clean malware (0.7832), its performance under attack significantly surpasses all other models. For instance, under 100-FSA, DeepTrust achieves a TPR of 0.1992, which is a relative improvement of 266\% over the second-best model, SecSVM. This substantial gain in robustness is achieved with only a minimal trade-off in the TNR (0.9902). In contrast, the baseline DREBIN model, despite its high TNR, sees its detection rate collapse to 0.0000 under the strongest attack, highlighting the critical security gap that DeepTrust effectively addresses.

\begin{table}[t]
    \centering
    \caption{Track 1 Competition Results. Best and second-best values are indicated in \textbf{bold} and \underline{underlined}, respectively. (*) Not submitted to the competition but aligned with prerequisites. (**) Not submitted to the competition and unaligned with requirements due to FPR$>$1\%}
    \begin{tabular}{p{0.25\columnwidth} p{0.05\columnwidth} p{0.05\columnwidth} p{0.06\columnwidth} p{0.06\columnwidth} p{0.06\columnwidth}}
    \toprule
    & \textbf{TNR} & \textbf{TPR no FSA} & \textbf{TPR 25-FSA} & \textbf{TPR 50-FSA} & \textbf{TPR 100-FSA} \\
    \midrule
    \textbf{DeepTrust} & .9902 & \textbf{.7832} & \textbf{.4328} & \textbf{.3384} & \textbf{.1992} \\
    \textbf{Baseline - SecSVM} \cite{10.1109/TDSC.2017.2700270} & .9956 & .7536 &
    .1496 & \underline{.1064} & \underline{.0544} \\
    \textbf{SVM-CB(b0.8,n100)} \cite{DBLP:conf/itasec/AngioniDPB22} & .9918 & .7536 & .0416 & .0328 & .0232 \\
    \textbf{SVM-CB(b0.2,n100)} & .9918 & .7536 & .0416 & .0328 & .0232 \\
    \textbf{C-PCT} \cite{ghiani2025regressionawarecontinuallearningandroid} & \underline{.9960} & .5960 & .0456 & .0120 & .0024 \\
    \textbf{Baseline - DREBIN} \cite{DBLP:conf/ndss/ArpSHGR14} & \textbf{.9964} & \underline{.7728} & .0048 & .0008 & .0000 \\
    \midrule
    \textbf{wAdvNet*} & .9956 & .7552 & \underline{.1704} & .0960 & .0432 \\
    \textbf{XGBoost*} \cite{10.1145/2939672.2939785} & .9988 & .7336 & .0016 & .0000 & .0000 \\
    \textbf{Random Forest*} \cite{tin_kam_ho_random_1995} & \underline{.9960} & .5440 & .0528 & .0336 & .0248 \\
    \textbf{SAdvNet**} & .9830 & .8016 & .4704 & .2656 & .1240 \\
    \bottomrule
    \end{tabular}
    \label{tab:track-1-classification}
\end{table}

\subsubsection{Track 2 - Adversarial Robustness to PSA}

In the problem-space attack scenario, DeepTrust reaffirms its effectiveness (Table \ref{tab:track-2-classification}). Although not an official competitor on this track, its evaluation shows remarkable resilience to direct APK manipulations. It achieves a detection rate of 0.7768 on adversarially modified samples, showing almost no degradation compared to its performance on clean data (0.7832). This result massively outperforms the next-best model, C-PCT, which scores 0.3584, representing a 117\% relative performance advantage for DeepTrust. This demonstrates that the defense mechanisms are effective against both powerful, hypothetical attackers, as feature-space attacks, and more practical, functionality-preserving attacks as in problem-space scenarios.

\begin{table}[t]
    \centering
    \caption{Track 2 Competition Results. Best and second-best values are indicated in \textbf{bold} and with an \underline{underline}, respectively. (*) Not submitted to the competition but aligned with prerequisites.}
    \begin{tabular}{p{0.25\columnwidth} p{0.06\columnwidth} p{0.06\columnwidth} p{0.06\columnwidth}}
    \toprule
    & \textbf{TNR} & \textbf{TPR no PSA} & \textbf{TPR 100-PSA} \\
    \midrule
    \textbf{C-PCT} & \underline{.9960} & .5920 & \underline{.3584} \\
    \textbf{Baseline - SecSVM} & .9956 & .7536 & .1168 \\
    \textbf{SVM-CB(b0.2,n100)} & .9918 & .7536 & .0488 \\
    \textbf{Baseline - DREBIN} & \textbf{.9964} & \underline{.7728} & .0376 \\
    \midrule
    \textbf{DeepTrust*} & .9902 & \textbf{.7832} & \textbf{.7768} \\
    \bottomrule
    \end{tabular}
    \label{tab:track-2-classification}
\end{table}

\subsubsection{Track 3 Results: Temporal Robustness to Data Drift}

Track 3 evaluates the models' resilience to concept drift over time. Here, DeepTrust demonstrates strong long-term stability, securing the second-place rank with an AUT-F1 score of 0.78192 (Table \ref{tab:track-3-classification}). It trails the top-performing DREBIN model by a negligible margin of just 1.36\%. This result is significant as it shows that the complex defensive architecture of DeepTrust does not compromise its ability to generalize to new, evolving data over time. Its performance is nearly on par with a simpler baseline, confirming its practicality for real-world deployment where both adversarial robustness and temporal stability are crucial.

\begin{table}[t!]
    \centering
    \caption{Track 3 Competition Results. Best and second-best values are indicated in \textbf{bold} and with an \underline{underline}, respectively.}

    \begin{tabular}{l c}
    \toprule
    & \textbf{AUT (F1 Score)} \\
    \midrule
    \textbf{Baseline - DREBIN} & \textbf{.7927} \\
    \textbf{DeepTrust} & \underline{.7819} \\
    \textbf{Baseline - SecSVM} & .7705 \\
    \textbf{SVM-CB(b0.2,n100)} & .7597 \\
    \textbf{SVM-CB(b0.8,n100)} & .7594 \\
    \textbf{DREBIN with features selection} & .6873 \\
    \textbf{C-PCT} & .6155 \\
    \bottomrule
    \end{tabular}
    \label{tab:track-3-classification}
\end{table}

\section{Conclusions}
\label{sec:conclusions}

This research presents DeepTrust, a novel metaheuristic that arranges deep neural networks in an ordered, multi-step sequence for robust Android malware detection. We empirically validated our central hypothesis: the efficacy of this approach hinges on maximizing the divergence of the learned representations among the internal models. Unlike traditional ensemble methods that aggregate predictions, our sequential decision-making forces an attacker to simultaneously deceive multiple models that induce fundamentally dissimilar data embeddings, frustrating the iterative perturbation process of evasion attacks. We achieved this representational diversity by training models with distinct regimes: a novel tabular Adversarial Training algorithm and label smoothing via distillation of a simpler algorithm.

The definitive validation of our method was its gold medal-winning performance in the IEEE SaTML'25 competition \cite{ramd-elsa-benchmark,ramd-elsa-competition}. Our submission, DeepTrust, achieved state-of-the-art results in feature-space evasion attacks (Track 1), outperforming the next-best competitor by up to 266\% in detection rate while strictly maintaining a false positive rate under 0.01. DeepTrust also demonstrated minimal degradation in problem-space attacks and strong temporal robustness to data drift. These results confirm that our multi-step architecture, driven by representational diversity, offers a superior and practical defense against sophisticated, real-world threats without compromising accuracy on non-perturbed data.

\subsection{Limitations and Future Work}
\label{sec:future_work}

Our work opens several avenues for future research. A primary limitation is that representational diversity was achieved as an emergent property of our training schemes rather than a directly measurable and optimizable one. Future work should formalize this property through representational similarity measures \cite{klabunde_similarity_2025}, like Centered Kernel Alignment (CKA) \cite{pmlr-v97-kornblith19a} or Canonical Correlation Analysis \cite{NEURIPS2018_a7a3d70c}, to quantitatively compare embeddings coming from different models. This would allow for the development of principled methods to induce representational diversity by design, perhaps through a unified loss function that co-optimizes for accuracy and inter-model dissimilarity. Building on frameworks like Negative Correlation Learning (NCL) \cite{LIU19991399} could provide a starting point.

\section*{Acknowledgments}
D.~Pulido-Cortázar was supported by a grant JAE-ICU 2024  funded by IIIA-CSIC (reference 04264).  D.~Gibert was supported by grant RYC2023-043607-I funded by MICIU/AEI/10.13039/501100011033 and FSE+. F.~Many\`a was supported by grant PID2022-139835NB-C21 funded by MCIN/AEI/10.13039/501100011033 and by ERDF, EU.

\bibliographystyle{unsrt}  
\bibliography{references}  

\appendix
\section{Evasion Attacks}
\label{app1}

The adversarial challenges in Tracks 1 and 2 are executed using two distinct attack methodologies, both based on genetic algorithms (GA) implemented with the DEAP (Distributed Evolutionary Algorithms in Python) framework \cite{fortin_deap_2012}. These attacks aim to modify malware samples to evade detection by minimizing the classifier's confidence score.

\subsection{Feature-Space Attack}
\label{sec:feature_space_attack}
The attack for Track 1 operates in the feature space, directly manipulating the DREBIN feature vectors of malware samples. It emulates an attacker with knowledge of the feature representation but interacts with the model in a black-box fashion, using only its output prediction scores to guide the attack. The simulation is implemented as a GA whose primary objective is to minimize the classifier's malware confidence score, thereby causing a misclassification from malware to benign. Each individual in the GA population is a fixed-length vector representing a set of manipulations, where a manipulation is encoded as an integer index corresponding to a feature in the classifier's known feature space. A positive index signifies the addition of a feature, while a negative index signifies its removal. The initial population of these manipulation vectors is derived from features present in a provided set of goodware samples, a strategy that seeds the algorithm with feature-addition candidates characteristic of benign applications. The feasibility of manipulations is predetermined. Certain DREBIN feature types can only be added, as defined in the attack implementation. The GA iteratively refines the population through an evolutionary process, where the fitness of each individual is the malware confidence score produced by the target classifier. A tournament selection process chooses individuals for the next generation. These individuals then undergo crossover, where two parents exchange a random subset of their manipulation indices, and mutation, where a manipulation is randomly replaced with a different, valid one from the overall manipulation space. The attack runs for a fixed number of iterations or stops early if a manipulation successfully evades the classifier.

\subsection{Problem-Space Attack}
\label{sec:problem_space_attack}
The attack for Track 2 operates in the problem space, simulating a more realistic black-box scenario by manipulating the APK files directly. The objective remains to modify an APK to minimize the classifier's malware score and achieve an evasion. In this GA, an individual is a vector of integer indices, where each index maps to a specific, pre-validated manipulation, such as injecting a permission or obfuscating an activity name. A key distinction of this attack is its two-stage process for defining valid manipulations. First, potential manipulations are generated by sourcing feature injections from a corpus of goodware APKs and identifying obfuscation candidates from the target malware itself. Second, a Manipulator tool built using the ObfuscAPK framework \cite{AONZO2020100403}, attempts to apply each candidate manipulation to a copy of the target APK. Only manipulations that can be applied without corrupting the APK or causing build errors are retained in the final, ``error-free" manipulation space available to the GA. This critical step ensures that all generated adversarial APKs are valid and functional. The subsequent fitness evaluation is computationally intensive, as it requires generating a new adversarial APK for each individual and running the full classification pipeline on it. This process is parallelized to manage the overhead. The evolutionary operators of tournament selection, crossover, and mutation are analogous to those in the feature-space attack but operate on the indices of these validated manipulations. The GA runs for a set number of iterations or until an evasion is successful, with the final output being the adversarially modified APK that achieved the lowest malware score.

\end{document}